\documentclass[12pt,prd,tightenlines,nofootinbib,showpacs,showkeys]{revtex4}
\usepackage{bm}
\usepackage{graphics}
\usepackage{epsfig}
\usepackage{amssymb}
\usepackage{amsmath}
\usepackage{amsthm}
\usepackage{amsfonts}
\usepackage{amssymb}
\usepackage{braket}
\usepackage{rotating}
\usepackage{setspace}
\usepackage{indentfirst}
\usepackage[T2A]{fontenc}
\usepackage[cp1251]{inputenc}
\usepackage[english]{babel}

\begin{document}
\title{Energy interval 3S-1S in muonic hydrogen}
\author{\firstname{A.~E.} \surname{Dorokhov}}
\affiliation{Joint Institute of Nuclear Research, BLTP, 141980, Moscow region,
Dubna, Russia}
\author{\firstname{R.N.} \surname{Faustov}}
\affiliation{Institute of Cybernetics and Informatics in Education, FRC CSC RAS, Moscow, Russia}
\author{\firstname{A.~P.} \surname{Martynenko}}
\author{\firstname{F.~A.} \surname{Martynenko}}
\affiliation{Samara National Research University, Moskovskoye Shosse 34, 443086, Samara, Russia}

\begin{abstract}
The energy interval $(3S-1S)$ in muonic hydrogen is calculated on the basis of 
quasipotential approach in quantum electrodynamics.
We take into account different corrections of orders $\alpha^3\div \alpha^6 $, 
which are determined by relativistic effects, the effects of vacuum polarization, 
nuclear structure and recoil, as well as combined corrections including the above. 
Nuclear structure effects
are expressed in terms of the charge radius of the proton in the case of one-photon interaction 
and the proton electromagnetic form factors in the case of two-photon exchange interaction.
The value of the energy interval $(3S-1S)$ can be used for a comparison 
with future experimental data and determining the proton charge radius with greater accuracy.
\end{abstract}



\maketitle

\section{Introduction}

At present, four complementary methods are used to obtain the charge radii of light nuclei: 
elastic scattering of electrons on nuclei, elastic scattering of muons on nuclei, 
spectroscopy of electron atoms,
high-precision laser spectroscopy in muonic atoms \cite{hill, paz, pineda1, bernauer}. 
Traditionally, elastic electron scattering was the first method to determine
internal structure of nuclei. Elastic scattering of leptons by a nucleus target is described 
by form factors included in the theoretical expression for the scattering cross section.
A proton or other light nucleus is a compound particle, and its size is determined by the 
charge radius
$r_{p}$. It is related to the slope of the electric form factor of the proton $G_{E}$ 
at $q^2= 0$. Since $G_{E} $ is a nonperturbative function of $q^2$, its slope 
should be extracted from experimental data. The most direct way to measure $r_{p}$ is to extract 
the electric form factor $G_{E}$ from lepton-proton scattering and determine its slope 
at $q^2$ = 0. Experimental data provide detailed information on the distribution 
of electric charge and magnetic moment inside the proton. They are used to determine
the absolute values of the proton charge $r_p$ and magnetic $r_M$ radii, usually
with a percentage accuracy or slightly better. At present, experiments are being carried out 
on the scattering of electrons and muons by protons.
By simultaneously determining the form factors for electron and muon scattering, 
these experiments will allow an accurate test of the universality of the lepton and,
thus contributing to the solution of the ''puzzle'' with the proton radius in the near future 
\cite{carlson,sgk2019}.

Atomic spectroscopy of hydrogen is an indirect way of determining the charge radius $ r_{p}$ 
of a proton from precision measurements of certain energy intervals. While electron scattering 
and spectroscopy of electron atoms were available
for a long time, muon spectroscopy became available only in 2010 due to the work of the 
CREMA collaboration. As a result of the first CREMA experiments in 2010, there was obtained
the value $ r_{p} = 0.84184 (67) $ fm, which was 10 times more accurate than all previous 
values from experiments with electronic systems. Moreover, this value was significantly less 
than the CODATA value, $ r_{p} = 0.8768 (69)$ fm. This difference is called the ''puzzle''
of the proton radius. It's safe to say that the decade since 2010 has been CREMA's decade. 
Let us list the main experimental results obtained during this time and published:

1. The measurement of transition frequency $(2S^{F=1}_{1/2}-2P^{F=2}_{3/2})$ in muonic
hydrogen in 2010 \cite{crema2010}.

2. The measurement of two transition frequencies $(2S^{F=1}_{1/2}-2P^{F=2}_{3/2})$ and
$(2S^{F=0}_{1/2}-2P^{F=1}_{3/2})$ in muonic hydrogen, the measurement of hyperfine structure of
$2S$ state in 2013 \cite{crema2013,crema}.

3. The measurement of three transition frequencies between energy levels 2P and 2S 
in muonic deuterium: 
$(2S^{F=3/2}_{1/2}-2P^{F=5/2}_{3/2})$,
$(2S^{F=1/2}_{1/2}-2P^{F=3/2}_{3/2})$, $(2S^{F=1/2}_{1/2}-2P^{F=1/2}_{3/2})$ 
\cite{crema2016}.

The performed studies with muonic hydrogen showed that there is a significant discrepancy 
in the values charge radius of the proton and deuteron, which are obtained
from experiments with electronic and muonic atoms. In the case of other light nuclei (helium), 
the results are preliminary and have so far been reported only at conferences.

During 2017-2019 different experimental results were obtained, both with electronic 
and muonic systems, which made it possible to extract the value of the charge radius 
of the proton. In \cite{beyer}, the frequency of the (2S-4P) transition in hydrogen 
was measured: $ \Delta \nu_{2S-4P} = 616520931626.8 (2.3) $ kHz, and the extracted value
$ r_{p} = 0.8335 (95) $ fm turned out to be in agreement with the CREMA
results \cite{crema}.

To investigate the puzzle of the proton radius, the PRad experiment was proposed 
in 2011 (E12-11-1062)
and was successfully carried out in 2016 at the Thomas Jefferson National Accelerator Facility 
with electron beams with energies of 1.1 and 2.2 GeV. In the experiment, the elastic scattering
cross sections $(e-p)$ were measured at unprecidentedly low values of the square of the momentum 
transfer with an accuracy of one percent. The value of the charge radius of the proton was 
$r_{p}=0.831 \pm 0.007 (stat) \pm 0.012 (syst)$ fm \cite{prad}, which is less than 
the average value of $ r_{p} $ from previous experiments on elastic scattering $(e-p)$, but is 
consistent with spectroscopic results for the muonic hydrogen atom within experimental 
uncertainties.

A new measurement of the Lamb shift in hydrogen $(n=2)$ was performed in \cite{bezginov}: 
$\Delta E^{Ls}=1057.8298(32)$ MHz
($909.8717(32)$ MHz). The value of the proton charge radius, which was 
obtained from this experiment, $r_{p}=0.833(10)$ fm, agrees with the spectroscopic 
data for muonic atoms.

To solve the problem of the proton charge radius, the MUSE collaboration is planning an 
experiment to simultaneously measure the cross sections for scattering of electrons and 
muons by protons \cite{gilman}. This experiment will make it possible to determine the 
charge radii of the proton independently in two reactions and test lepton universality 
with an accuracy of an order of magnitude superior to previous scattering experiments.

It should be noted that in the recent experiment \cite{fleurbaey}, a new measurement 
of the frequency of the two-photon transition $(1S-3S)$ in hydrogen was carried out 
with a relative error $ 9 \cdot 10^{- 13} $: $\Delta \nu^{2017}_{1S-3S}= 
2922743278671.0(4.9)$ kHz. The value of the charge radius extracted from this experimental 
result, $r_{p}=0.877(13)$ fm, is in good agreement with the value recommended 
by the CODATA \cite{codata}.

The experimental accuracy of measuring the energy intervals between the $S$-levels of the hydrogen 
atom, muonium is very high and continues to grow. There is a clear prospect of using such 
intervals to search for the effects of New Physics beyond the Standard Model. Spectroscopy 
of purely lepton systems, as well as light muonic atoms, can help in assessing the possible 
manifestations of spin-dependent and spin-independent forces of dark matter. The result of the 
experiment \cite{fleurbaey} in electron hydrogen posed, in our opinion, two problems. 
First, it is necessary to re-analyze the theoretical calculation of the various contributions 
to the $(3S-1S)$ interval in hydrogen in order to obtain the total theoretical value for the $(3S-1S)$ 
transition frequency and extract the proton charge radius (see the recent work 
\cite{yerokhin2018}). The situation with the experiment \cite{fleurbaey}, 
which gives a different magnitude for
the charge radius of the proton in comparison with \cite{beyer, prad, bezginov} requires 
a new consideration. The discrepancy in the proton charge radii of 0.03 fm can be due to 
contributions of order 100 kHz. Second, it is useful to have a precise theoretical 
calculation of the $(3S-1S)$ energy interval in muonic hydrogen as a guideline for possible future 
experiments with muonic hydrogen. This work is aimed at solving the second problem.

\section{Effects of vacuum polarization in one-photon interaction}

The main parameters of nuclei, such as the charge radius, quadrupole moment, magnetic octupole 
moment, etc. are known primarily from experiments on
scattering of leptons on nuclei, which have been carried out for many years. The accuracy 
of their determination is not very high. Another approach to the study of these parameters
is related to the measurement of various spectroscopic 
intervals in electronic or muonic atoms and ions with such nuclei.
We are considering one of such basic intervals $(3S-1S)$, 
which is measured for the electron hydrogen atom with very high accuracy. One of the leading 
contributions to this energy interval is determined by the proton charge radius, and, therefore, 
the charge radius can be extracted from the corresponding experimental data.

Our approach to the precision calculation of the energy range $(3S-1S)$ is based 
on the quasipotential method in quantum electrodynamics \cite{apm2005,apm2015,apm2019}.
The two-particle bound state is described by the Schr\"odinger equation, and the main 
contribution to the particle interaction operator is determined by the Breit Hamiltonian. 
A number of important results in the study of the energy levels of muonic atoms have been 
obtained in \cite{friar,borie,kp1996} (see \cite{egs} for other references). 
The main contribution to the fine structure of the S-wave spectrum of hydrogen-like atoms 
consisting of particles with masses $m_1$ (the muon mass), $m_2$ (the proton mass) can be 
represented with an accuracy of $O((Z \alpha)^6)$ ($\mu$ is the reduced mass) 
in the form \cite{egs}:
\begin{equation}
\label{eq1}
E_n=m_1+m_2-\frac{\mu(Z\alpha)^2}{2n^2}-\frac{\mu(Z\alpha)^4}{2n^3}\left[1-\frac{3}{4n}+
\frac{\mu^2}{4m_1m_2n}\right]-\frac{m_1(Z\alpha)^6}{16n^6}(2n^3+6n^2-12n+5).
\end{equation}
This formula correctly takes into account the recoil correction $m_1^2/m_2^2 (Z\alpha)^4$ 
for nuclei with spin 1/2 \cite{egs} (see section 4). Recoil effects of order 
$(Z\alpha)^6$ are not taken into account in formula \eqref{eq1} and are discussed in Section 5.

To extract the charge radius of a proton with high accuracy from the measurement of some 
spectroscopic interval, it is necessary to calculate corrections of a high order of smallness 
in terms of the fine structure constant and the particle mass ratio. Numerical values 
of the same effects presented in the Feynman diagrams differ significantly in the case 
of electronic and muonic hydrogen. 
In what follows, we consider the calculation of contributions to the $(3S-1S)$ interval in order 
of importance for muonic hydrogen. Numerical values of contributions are shown in
separate lines of Table~\ref{tb1}. In the second column of Table~\ref{tb1} we
indicate the order of the considered contribution for muonic hydrogen.

An important class of corrections to energy levels are corrections for the vacuum polarization (VP). 
Although their value decreases with an increase in the number of loops in the polarization 
operator, it is necessary to take into account contributions up to three loops inclusively 
to achieve a high calculation accuracy. The one-loop vacuum polarization leads 
to a modification of the Coulomb potential and is determined in the coordinate representation 
by the following expression (the subscript $vp$ denotes here and below the electronic 
polarization of the vacuum, and the superscript $C$ the contribution of the Coulomb 
interaction):
\begin{equation}
\label{eq2}
V^C_{vp}(r)=\frac{\alpha}{3\pi}\int_1^\infty d\xi \rho(\xi)
\left(-\frac{Z\alpha}{r}e^{-2m_e\xi r}\right),~~~
\rho(\xi)=\frac{\sqrt{\xi^2-1}(2\xi^2+1)}{\xi^4}.
\end{equation}

\begin{figure}[htbp]
\centering
\includegraphics[scale=0.7]{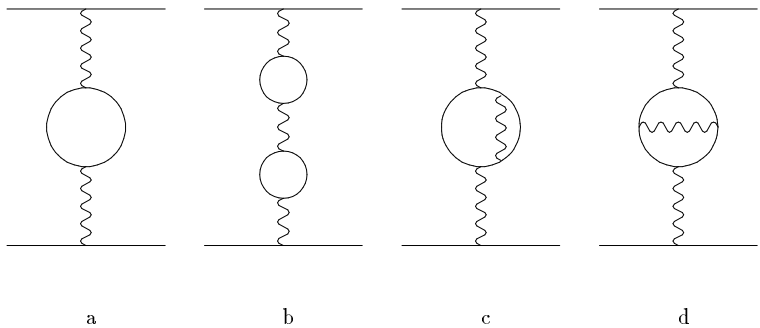}
\caption{Effects of one-loop and two-loop vacuum polarization in one-photon
interaction}
\label{fig11}
\end{figure}

In the first order of the perturbation theory (PT), it is necessary to calculate the matrix 
elements of the operator \eqref{eq2} from the wave functions of $1S$ and $3S$ states, which 
have the form:
\begin{equation}
\label{eq3a}
\psi_{100}(r)=\frac{W^{3/2}}{\sqrt{\pi}}e^{-Wr},~~~W=\mu Z\alpha,
\end{equation}
\begin{displaymath}
\psi_{300}(r)=\frac{W^{3/2}}{3\sqrt{3\pi}}e^{-Wr/3}\left(1-\frac{2}{3}Wr+\frac{2}{27}(Wr)^2\right).
\end{displaymath}

\begin{figure}[htbp]
\centering
\includegraphics[scale=0.8]{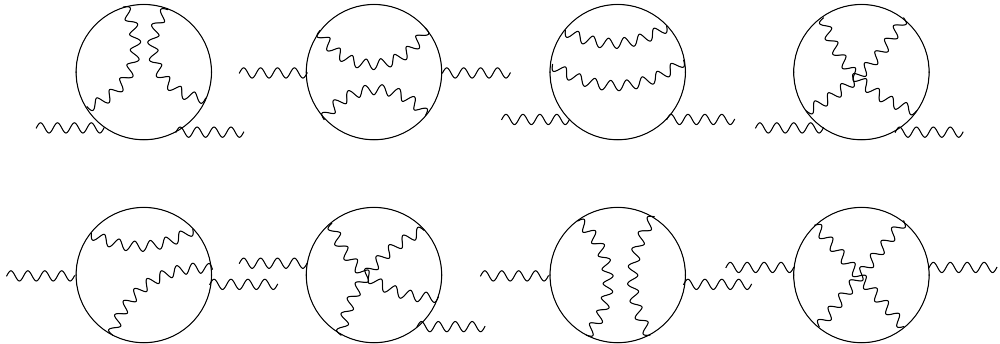}
\caption{Effects of three-loop vacuum polarization 
with one fermionic cycle in one-photon interaction}
\label{fig22}
\end{figure}

\begin{figure}[htbp]
\centering
\includegraphics[scale=1.0]{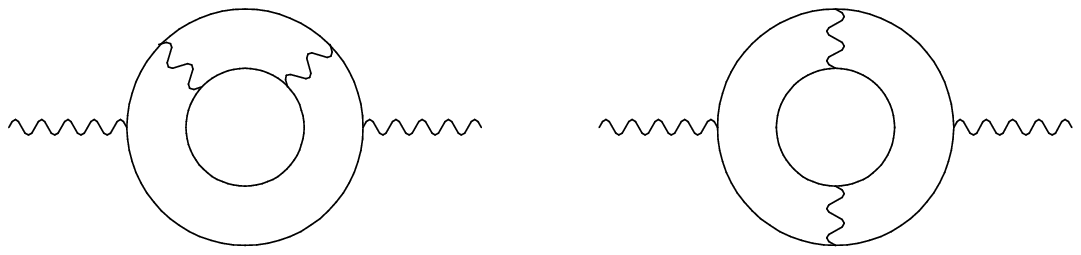}
\caption{Effects of three-loop vacuum polarization with two fermionic cycles 
in one-photon interaction}
\label{fig33}
\end{figure}

\begin{figure}[htbp]
\centering
\includegraphics[scale=0.7]{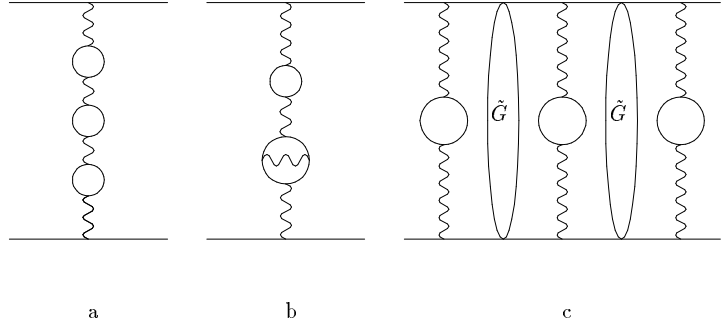}
\caption{Effects of three-loop vacuum polarization in one-photon interaction and third-order 
perturbation theory}
\label{fig33_1}
\end{figure}

Analytical calculation of matrix elements gives the following shifts of energy levels 
$1S$ and $3S$ ($b_1=m_e/W$, $W=\mu Z\alpha$, $Z=1$ is the charge of the proton):
\begin{equation}
\label{eq3}
\Delta E_{vp}(1S)=-\frac{4\mu(Z\alpha)^2\alpha}{3\pi}
\frac{\sqrt{{b_1}^2-1} \left(12 \pi  {b_1}^3-24 {b_1}^2+9 \pi  
{b_1}-22\right)-6 \left(4 {b_1}^4+{b_1}^2-2\right) \sec^{-1}({b_1})}
{6 \sqrt{{b_1}^2-1}},
\end{equation}
\begin{equation}
\label{eq4}
\Delta E_{vp}(3S)=-\frac{4\mu\alpha(Z\alpha)^2}{81\pi}\frac{1}
{16 \bigl(9 b_1^2-1\bigr)^{9/2}}\Bigl\{
\sqrt{9 b_1^2-1} \bigl(9 b_1 \bigl(-7243344 b_1^9+3006396 b_1^7-447768 b_1^5+
\end{equation}
\begin{displaymath}
27999 b_1^3+18 \pi  \bigl(1-9
b_1^2\bigr)^4 \bigl(92 b_1^2+1\bigr)-224 b_1\bigr)-64\bigr)-3 i 
\bigl(81 b_1^2 \bigl(27 \bigl(\bigl(108 b_1^2
\bigl(828 b_1^4-405 b_1^2+76\bigr)-
\end{displaymath}
\begin{displaymath}
703\bigr) b_1^2+26\bigr) b_1^2+4\bigr)-8\bigr) 
\ln\Bigl(\frac{3 ib_1}{\sqrt{9 b_1^2-1}-i}\Bigr)\Bigr\}.
\end{displaymath}
It is useful to note that the $1/b_1=\mu Z\alpha/m_e=1.356$, so the parameter $1/b_1$ 
cannot be used as an expansion parameter, since it is not small.
The corresponding numerical value of these contributions for the $(3S-1S)$ interval 
is presented in Table~\ref{tb1} (line 2). It is written for definiteness with 
an accuracy of up to 4 digits after the decimal point, since the errors connected 
with errors in the determination of fundamental physical constants
are significantly less. The expressions \eqref{eq3}-\eqref{eq4} can also be used to numerically 
estimate the contribution of muon vacuum polarization in muonic hydrogen, replacing 
the electron mass with the muon mass. In this case, numerical value of the contribution 
decreases sharply, which is connected with an increase in the general order of such 
a contribution due to the factor $\alpha^2$ (see line 3 in Table~\ref{tb1}). 
Numerical value of this correction is somewhat different from that which is obtained
using an analytical expression in leading order 
$ \Delta E_{mvp}(3S-1S)=104\alpha (Z\alpha)^4\mu^3/405\pi m_1^2=0.1298$ meV.
It appears in \eqref{eq3}-\eqref{eq4} due to corrections
of higher order and must be taken into account if we want to use this interval for more accurate 
obtaining of the proton charge radius. The operator \eqref{eq2} also makes contributions in higher 
orders of perturbation theory, which are considered below. In one-photon interaction
there are also contributions of two-loop and three-loop vacuum polarization 
(see Fig.~\ref{fig11}, Fig.~\ref{fig22}, Fig.~\ref{fig33}, Fig.~\ref{fig33_1}).

In the case of contributions from the fourth-order polarization operator 
(Fig.~\ref{fig11}), one can construct the interaction potential of particles 
and the shift of energy levels in integral form using the replacement 
in the photon propagator in the momentum representation \cite{cs}:
\begin{equation}
\label{k2}
\frac{1}{k^2}\to\frac{2}{3}\left(\frac{\alpha}{\pi}\right)^2\int_0^1\frac{f(v)dv}
{4m_e^2+k^2(1-v^2)},
\end{equation}
\begin{displaymath}
f(v)=v\Bigl\{(3-v^2)(1+v^2)\left[Li_2\left(-\frac{1-v}{1+v}\right)+2Li_2
\left(\frac{1-v}{1+v}\right)+\frac{3}{2}\ln\frac{1+v}{1-v}\ln\frac{1+v}{2}-
\ln\frac{1+v}{1-v}\ln v\right]
\end{displaymath}
\begin{displaymath}
+\left[\frac{11}{16}(3-v^2)(1+v^2)+\frac{v^4}{4}\right]\ln\frac{1+v}{1-v}+
\left[\frac{3}{2}v(3-v^2)\ln\frac{1-v^2}{4}-2v(3-v^2)\ln v\right]+
\frac{3}{8}v(5-3v^2)\Bigr\},
\end{displaymath}
where $Li_2(z) $ is the Euler dilogarithm. Then, in the coordinate representation, the particle 
interaction operator takes the form convenient for the subsequent calculation 
of the energy shift:
\begin{equation}
\label{cs}
\Delta V^C_{1\gamma,2-loop~vp}(r)=-\frac{2Z\alpha}{3r}\left(\frac{\alpha}{\pi}\right)^2
\int_0^1\frac{f(v)dv}{1-v^2}e^{-\frac{2m_er}{\sqrt{1-v^2}}}.
\end{equation}
The numerical contribution \eqref{cs} is included in the Table~\ref{tb1} together 
with another contribution of the two-loop vacuum polarization with two successive loops, 
which is denoted below by the subscript $vp-vp$. In the momentum representation, 
the corresponding particle interaction potential has the form:
\begin{equation}
\label{vp-vp}
V^C_{vp-vp}(k^2)=-4\pi(Z\alpha)\frac{\alpha^2}{9\pi^2}\int_1^\infty\rho(\xi)d\xi\int_1^\infty
\rho(\eta)d\eta\frac{k^2}{(k^2+4m_e^2\xi^2)(k^2+4m_e^2\eta^2)}=
\end{equation}
\begin{displaymath}
=-\frac{2\alpha^2(Z\alpha)}{9\pi}\int_1^\infty\rho(\xi)d\xi\int_1^\infty\rho(\eta)d\eta
\Bigl\{\frac{1}{k^2+4m_e^2\xi^2}+\frac{1}{k^2+4m_e^2\eta^2}-\frac{(\xi^2+\eta^2)}{(\eta^2-\xi^2)}
\end{displaymath}
\begin{displaymath}
\Bigl[
\frac{1}{k^2+4m_e^2\xi^2}-\frac{1}{k^2+4m_e^2\eta^2}\Bigr]\Bigr\}.
\end{displaymath}
After the Fourier transform, \eqref{vp-vp} takes the form of a superposition of the Yukawa 
potentials, distributed with a certain density:
\begin{equation}
\label{vp-vp-1}
V^C_{1\gamma,vp-vp}(r)=\frac{\alpha^2}{9\pi^2}\int_1^\infty\rho(\xi)d\xi\int_1^\infty\rho(\eta)d\eta
\left(-\frac{Z\alpha}{r}\right)\frac{1}{(\xi^2-\eta^2)}(\xi^2e^{-2m_e\xi r}-\eta^2 e^{-2m_e\eta r}).
\end{equation}
When calculating the matrix elements \eqref{cs}, \eqref{vp-vp-1}, the integration 
over the particle coordinates is performed analytically, and the subsequent integration over 
the spectral parameters is numerical. The contributions \eqref{cs}, \eqref{vp-vp-1} 
are of order $ \alpha^2 (Z \alpha)^2 $ for muonic hydrogen and are numerically large 
(see line 6 in the Table), so it is necessary to consider
corrections of the next order in $\alpha$. In the case of muon vacuum polarization, 
the two-loop contributions \eqref{cs}, \eqref {vp-vp-1} equal to 0.0011 meV has the order 
$ \alpha^2 (Z \alpha)^4 $ and is determined by the well-known analytical formula 
(see Section 4 \cite{egs}). We have also included it in the full result of line 6.

Turning to the contributions of the 6th order polarization operator (in this case, 
the number 6 means the number of interaction vertices in the vacuum loop), we note 
that they were studied in the case of
the Lamb shift (2P-2S) in \cite{kn,kn1}. It is convenient to divide this contribution 
into two parts with one (1f) and two (2f) fermionic cycles. It is useful to note that in 
\cite{kl} a general parametric formula was obtained for the contributions in 
Fig.~\ref{fig22}, Fig.~\ref{fig33}, but it is difficult to use it to 
obtain numerical estimates due to the remaining multiple integrals over the Feynman 
and spectral parameters, as well as the renormalization procedure. A more convenient 
formula for practical use was obtained in \cite{baikov,kn} for the contribution 
of 8 diagrams:
\begin{equation}
\label{p31}
\Pi_3^{(1)}(z)=\tilde\Pi_3^{(1)}(z)-4\Pi_2(z)-(1-z)G(z)\left(\frac{9}{4}G(z)+\frac{31}{16}+
\frac{229}{32z}+\frac{229}{32z}+\frac{173}{96}\right),
\end{equation}
\begin{displaymath}
\tilde\Pi_3^{(1)}(z)=\tilde\Pi_3^{(1)}(z)(-\infty)+\frac{(1+\omega)^2}{(1-\omega)}
\frac{(\tilde a_0+\tilde a_1\omega+\tilde a_2\omega
+\tilde a_3\omega)}{(\tilde b_0+\tilde b_1\omega+\tilde b_2\omega+\tilde b_3\omega)},
\end{displaymath}
and the explicit form of the functions involved and their asymptotic values 
can be found in \cite{baikov, kn}. The Pade approximation coefficients 
$\tilde a_i$, $\tilde b_i$ are written out in \cite{kn1}.
The general formula for the contribution of the polarization operator \eqref{p31} 
(the contribution of 8 diagrams in Fig.~\ref{fig22}) to the shift $(3S-1S)$ 
has the form:
\begin{equation}
\label{p31a}
\Delta E_{1\gamma}^{1f}(3S-1S)=\frac{64}{\pi}\left(\frac{\alpha}{\pi}\right)^3\mu(Z\alpha)^2
\int_0^\infty\frac{s^2\rho_1(s)}
{(4+9s^2)^6(4+s^2)^2}\Pi^{(1)}_{3}\left(\frac{s^2}{4b_1^2}\right),
\end{equation}
\begin{displaymath}
\rho_1(s)=(69632+s^2(-71936+3 s^2 (239360+81s^2(3296+2688s^2+1053 s^4)))),
\end{displaymath}
and the numerical value of the contribution to the $(3S-1S)$ interval in muonic hydrogen is
\begin{equation}
\label{ep31}
\Delta E^{1f}_{1\gamma}(3S-1S)=0.0169~meV.
\end{equation}

Diagrams with two fermionic cycles in Fig.~\ref{fig33}
can be considered as corrections to the mass and vertex operators. In this case, 
the dispersion relation can be used for the second-order polarization operator. 
The effective propagator of a virtual photon in these diagrams can be represented as
\begin{equation}
\label{prop}
\frac{-i}{k^2+i0}\left(g_{\mu\lambda}-\xi\frac{k_\mu k_\lambda}{k^2}\right)
\Pi_{\lambda\sigma}(k^2)
\frac{-i}{k^2+i0}\left(g_{\nu\sigma}-\xi\frac{k_\nu k_\sigma}{k^2}\right),
\end{equation}
which, taking into account the transverse character of the polarization operator 
$ \Pi_{\lambda \sigma}(k^2) $, means that the result is gauge invariant. 
Taking into account such contributions is important for achieving high accuracy 
in calculating the anomalous magnetic moment of the lepton \cite{kataev,kataev1}.
The imaginary part of the polarization operator in Fig.~\ref{fig33} was initially 
represented in the form of a two-dimensional spectral integral \cite{hoang}, 
and then in an analytical form in \cite{hoang1}, and the final formula is rather 
cumbersome. In our calculations, we use the last representation of the polarization 
operator from \cite{hoang1}, and the calculation formula for the contribution is:
\begin{equation}
\label{p31b}
\Delta E_{1\gamma}^{2f}(3S-1S)=-\frac{64}{\pi}\left(\frac{\alpha}{\pi}\right)^3\mu(Z\alpha)^2
\int_0^\infty\frac{s^2\rho_1(s) ds}{(4+s^2)^2(4+9s^2)^6}\int_1^\infty
\frac{\rho_{2f}(t,s)dt}{t\left(t+\frac{s^2}{4\beta^2}\right)},
\end{equation}
where  two-fermion spectral density $\rho_{2f}(s,t)$ is taken from \cite{hoang1}, 
and the numerical value \eqref{p31b} for muonic hydrogen (-0.0158~meV) differs 
in sign from \eqref{p31a}. The sum of corrections \eqref{p31a} and \eqref{p31b} 
is equal to 0.0246 meV.

The remaining three-loop contributions in Fig.~\ref{fig33} with successive 
loops can be calculated in the same way as the two-loop ones in Fig.~\ref{fig11}, 
by constructing the interaction potentials as it was done in 
\eqref{vp-vp}-\eqref{vp-vp-1}. General expressions for these potentials 
in the coordinate representation are \cite{apm2005,apm2007}:
\begin{equation}
\label{vp-vp-vp}
V^C_{vp-vp-vp}(r)=-\frac{Z\alpha}{r}\frac{\alpha^3}{(3\pi)^3}\int_1^\infty
\rho(\xi)d\xi\int_1^\infty\rho(\eta) d\eta\int_1^\infty\rho(\zeta)d\zeta\times
\end{equation}
\begin{displaymath}
\times\left[e^{-2m_e\zeta r}\frac{\zeta^4}{(\xi^2-\zeta^2)(\eta^2-\zeta^2)}
+e^{-2m_e\xi r}\frac{\xi^4}{(\zeta^2-\xi^2)(\eta^2-\xi^2)}+
e^{-2m_e\eta r}\frac{\eta^4}{(\xi^2-\eta^2)(\zeta^2-\eta^2)}\right],
\end{displaymath}
\begin{equation}
\label{vp-2loopvp}
V^C_{vp-2-loop~vP}=-\frac{4\mu\alpha^3(Z\alpha)}{9\pi^3r}\int_1^\infty
\rho(\xi)d\xi\int_1^\infty\frac{f(\eta)d\eta}{\eta}
\left[e^{-2m_e\eta r}\frac{\eta^2}{\eta^2-\xi^2}-e^{-2m_e\xi r}\frac{\xi^2}
{\eta^2-\xi^2}\right].
\end{equation}
Corrections in the energy spectrum corresponding to these interactions are presented 
in integral form over three spectral parameters and calculated numerically.
The total numerical value of the three-loop contribution of the vacuum polarization 
from the $1\gamma $-interaction is presented in the Table~\ref{tb1} in a separate 
line 7. In the case of muon vacuum polarization, the contribution of the third-order 
polarization operator obtained by formula (58) \cite{egs} is negligible small.

\section{Effects of vacuum polarization and relativistic corrections}

The Breit potential contributes to the energy of $S$ -states in the leading order 
$(Z\alpha)^4$. The effect of vacuum polarization leads to a change not only 
in the Coulomb potential, but also in other terms in the Breit potential, which will 
contribute to the energy spectrum of order $ \alpha(Z\alpha)^4 $. 
This order of contribution suggests that the numerical values of the corrections 
can be significant. The modification of the Breit potential due to the one-loop 
vacuum polarization is determined in the case of S-states by the following terms 
(the superscript $B$ denotes the Breit potential) \cite{kp1996,apm2007,apm2016}:
\begin{equation}
\label{rel1}
\Delta V^B_{vp}(r)=\frac{\alpha}{3\pi}\int_1^\infty\rho(\xi)d\xi\sum_{i=1}^3
\Delta V_{i,vp}^B(r),
\end{equation}
\begin{equation}
\label{rel2}
\Delta V_{1,vp}^B=\frac{Z\alpha}{8}\left(\frac{1}{m_1^2}+\frac{1}{m_2^2}\right)
\left[4\pi\delta({\bf r})-\frac{4m_e^2\xi^2}{r}e^{-2m_e\xi r}\right],
\end{equation}
\begin{equation}
\label{rel3}
\Delta V_{2,vp}^B=-\frac{Z\alpha m_e^2\xi^2}{m_1m_2r}e^{-2m_e\xi
r}(1- m_e\xi r),
\end{equation}
\begin{equation}
\label{rel4}
\Delta V_{3,vp}^B=-\frac{Z\alpha}{2m_1m_2}p_i\frac{e^{-2m_e\xi r}}{r}
\left[\delta_{ij}+\frac{r_ir_j}{r^2}(1+2m_e\xi r)\right]p_j.
\end{equation}
The largest numerical contribution (more than 80 $\% $) comes from the term 
$ \Delta V_{1,vp}^B $, whose matrix elements are calculated analytically for 
$1S$ and $3S $ states:
\begin{equation}
\label{breitvp1}
\Delta E^B_{1,vp}(1S)=\frac{\alpha(Z\alpha)^4\mu^3}{18\pi}(\frac{1}{m_1^2}+\frac{1}{m_2^2})
\Biggl[\left(1+6b_1^2-3b_1^3\pi\right)
+\frac{1}{\sqrt{1-b_1^2}}(6-3b_1^2+6b_1^4)\ln \frac{1+\sqrt{1-b_1^2}}{b_1}
\Biggr],
\end{equation}
\begin{equation}
\label{breitvp2}
\Delta E^B_{1,vp}(3S)=\frac{\alpha(Z\alpha)^4\mu^3}{1944\pi(1-9b_1^2)^4}
(\frac{1}{m_1^2}+\frac{1}{m_2^2})
\Biggl[9 \bigl(9 b_1 \bigl(3 b_1 \bigl(6 \left(972 b_1^4-414 b_1^2-5\right)
b_1^2+19\bigr)-
\end{equation}
\begin{displaymath}
4 \pi  \left(1-9 b_1^2\right)^4\bigr)-116\bigr) b_1^2+32+
\frac{1}{\left(1-9 b_1^2\right)^{1/2}}
\Bigl[3\bigl(81 \bigl(9 b_1^2 \bigl(\bigl(162 b_1^2 \bigl(36 b_1^4-18b_1^2+5\bigr)-89\bigr)
b_1^2+
\end{displaymath}
\begin{displaymath}
10\bigr)-4\bigr) b_1^2+8\bigr) \ln \left(\frac{3 b_1 
\sqrt{1-9 b_1^2}}{-1+9 b_1^2+\sqrt{1-9b_1^2}}\right)\Bigr]\Biggr].
\end{displaymath}
The total contribution of relativistic corrections taking into account the one-loop 
vacuum polarization is presented in line 8 of Table~\ref{tb1}. In order to increase 
the accuracy of the calculation
we also take into account the contribution of relativistic corrections with effects 
of two-loop vacuum polarization of order $ \alpha^2(Z\alpha)^4 $. 
The main term in the interaction potential is:
\begin{equation}
\label{rel22}
\Delta V_{1,2loop-vp}^B=\frac{\alpha^2(Z\alpha)}{12\pi^2}\left(\frac{1}{m_1^2}+\frac{1}{m_2^2}\right)
\int_0^1\frac{f(v)dv}{1-v^2}
\left[4\pi\delta({\bf r})-\frac{4m_e^2}{r(1-v^2)}e^{-\frac{2m_e r}{\sqrt{1-v^2}}}\right],
\end{equation}
and the calculation of matrix elements is carried out similarly to 
\eqref{breitvp1}-\eqref{breitvp2}.

In the second order of perturbation theory, there are a number of contributions 
in which the potentials $\Delta V_{vp}^C $, $ \Delta V^B $ (Breit potential), 
$ \Delta V^B_{vp} $ are considered as perturbation operators (the subscript 
$sopt$ denotes the second-order PT contribution):
\begin{equation}
\label{pt1}
\Delta E_{sopt}^{B,vp}=<\psi|\Delta V^C_{vp}\tilde G\Delta V^C_{vp}|\psi>+
2<\psi|\Delta V^B\tilde G\Delta V^C_{vp}|\psi>+
\end{equation}
\begin{displaymath}
2<\psi|\Delta V^B_{vp}\tilde G\Delta V^C_{vp}|\psi>+
2<\psi|\Delta V^B\tilde G\Delta V^C_{vp,vp}|\psi>,
\end{displaymath}
\begin{equation}
\label{breit}
\Delta V^B=-\frac{{\bf p}^4}{8m_1^3}-\frac{{\bf p}^4}{8m_2^3}+\frac{\pi Z\alpha}{2}\left(\frac{1}{m_1^2}+
\frac{1}{m_2^2}\right)\delta({\bf r})
-\frac{Z\alpha}{2m_1m_2r}\left({\bf p}^2+\frac{{\bf r}({\bf rp}){\bf p}}
{r^2}\right).
\end{equation}

\begin{figure}[htbp]
\centering
\includegraphics[scale=0.7]{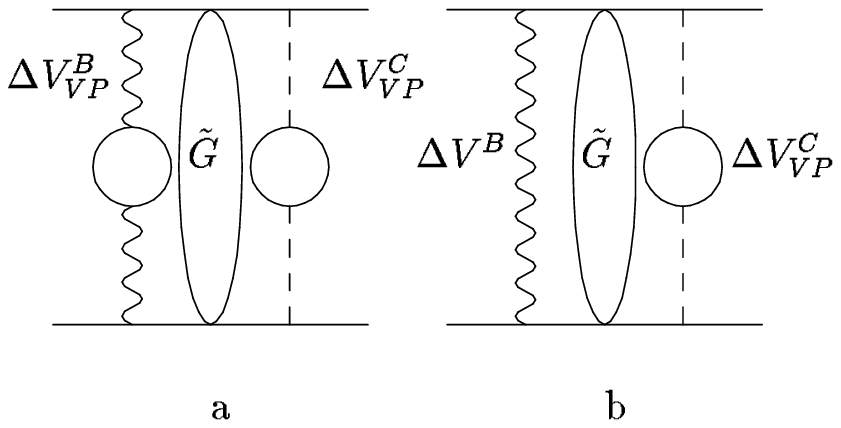}
\includegraphics[scale=0.7]{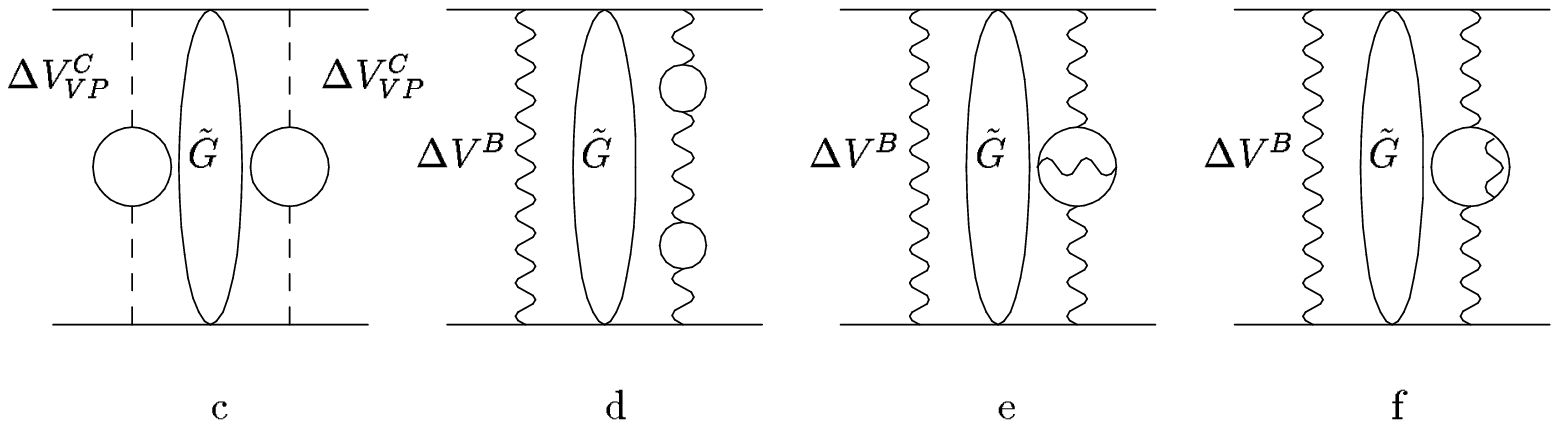}
\caption{Effects of one-loop and two-loop vacuum polarization in second order perturbation 
theory. $\tilde G$ is the reduced Coulomb Green's function.
The dashed line represents the Coulomb photon. The wavy line denotes the Breit potential 
(relativistic correction).}
\label{fig1}
\end{figure}

Such contributions, presented for a clarity in the diagrams in Fig.~\ref{fig1}, 
can be considered as one-loop and two-loop vacuum polarization corrections with 
account for relativistic effects. The reduced Coulomb Green's function of 1S and 2S 
states has the well-known form \cite{hameka}. To derive the Green's function 
for the $3S$ state, we used the general expression for the Coulomb Green's function 
\cite{veselov,sgk1996} in terms of the product of Whittaker's functions. 
After subtracting the pole term, the following expression is obtained 
for the reduced Green's function of the $3S$ state:
\begin{equation}
\label{green3S}
G_{3S}({\bf r}_1,{\bf r}_2)=-\frac{Z\alpha\mu^2}{13122\pi r_1 r_2}e^{-\frac{1}{3}(x_1+x_2)}g_{3S}(x_1,x_2),
\end{equation}
\begin{displaymath}
g_{3S}(x_1,x_2)=18 x_< (2 (x_<-9) x_<+27) (2 (x_>-9) x_>+27) x_> 
\Bigl(\text{Ei}\Bigl(\frac{2 x_<}{3}\Bigr)-\ln(x_<)-\ln \Bigl(\frac{4 x_>}{9}\Bigr)\Bigr)-
\end{displaymath}
\begin{displaymath}
4 x_< (2 (x_<-9) x_<+27) x_>^4+2 
\bigl(-27 e^{\frac{2}{3}x_<} (x_< (2 x_<-15)+9)+x_< (-36 \gamma  (2 (x_<-9) x_<+27)-
\end{displaymath}
\begin{displaymath}
2 x_< (x_< (2x_<-135)+891)+1701)+243\bigr) x_>^3+18 \bigl(27 e^{\frac{2}{3} x_<} 
(x_< (2 x_<-15)+9)+
\end{displaymath}
\begin{displaymath}
x_< (36 \gamma  
(2(x_<-9) x_<+27)+2 x_< (x_< (2 x_<-99)+567)-729)-243\bigr) x_>^2+
\end{displaymath}
\begin{displaymath}
27 \bigl(-27 e^{\frac{2}{3} x_<}
(x_< (2 x_<-15)+9)+x_< (-2 (x_<-27) x_<(2 x_<-9)-36 \gamma(2 (x_<-9)
x_<+
\end{displaymath}
\begin{displaymath}
27)-243)+243\bigr) x_>+243 x_< (2 (x_<-9) x_<+27),
\end{displaymath}
where $x_<=min(x_1,x_2)$,  $x_>=max(x_1,x_2)$, $x_i=Wr_i$, $\gamma$ is the Euler constant.

When calculating corrections in the second order PT, it is also necessary 
to know the reduced Green's function with one zero argument, which is obtained 
from \eqref{green3S} by expanding at $r_2=0$ in the form:
\begin{equation}
\label{green3SS}
G_{3S}({\bf r})=\frac{Z\alpha\mu^2}{3\pi x}e^{-\frac{1}{3}x}
\Bigl[4 x^4-144 x^3+648 x^2+18 \gamma  \left(2 x^2-18 x+27\right) x+
\end{equation}
\begin{displaymath}
18 \left(2 x^2-18 x+27\right) x \ln \left(\frac{2 x}{3}\right)-243\Bigr],
\end{displaymath}
where the dimensionless variable $x=Wr$.

Among the amplitudes in Fig.~\ref{fig1}, the largest contribution of order 
$\alpha^2(Z\alpha)^2$ is given by the amplitude (c), which contains two Coulomb potentials 
corrected for the vacuum polarization. An integration over coordinates 
can also be carried out analytically, and over spectral parameters numerically. Since, 
after the integration over coordinates, the result has a cumbersome form, we present here the initial 
integral expression for this correction and its numerical value in the shift $(3S-1S)$:
\begin{equation}
\label{vpvp1s}
\Delta E^{vp,vp}_{sopt}(1S)=-\frac{16\mu\alpha^2(Z\alpha)^2}{9\pi^2}
\int_1^\infty\rho(\xi)d\xi\int_1^\infty\rho(\eta)d\eta\times
\end{equation}
\begin{displaymath}
\int_0^\infty x_1 e^{-x_1\left(1-\frac{2m_e\xi}{W}
\right)}dx_1\int_0^\infty x_2 e^{-x_2\left(1-\frac{2m_e\eta}{W}\right)}g_{1S}(x_1,x_2)dx_2,
\end{displaymath}
\begin{equation}
\label{vpvp2s}
\Delta E^{vp,vp}_{sopt}(3S)=-\frac{8\mu\alpha^2(Z\alpha)^2}{1594323\pi^2}
\int_1^\infty\rho(\xi)d\xi\int_1^\infty\rho(\eta)d\eta\times
\end{equation}
\begin{displaymath}
\int_0^\infty\left(1-\frac{2x_1}{3}+\frac{2x_1^2}{27}\right)e^{-x_1\left(2/3-\frac{2m_e\xi}{W}
\right)}dx_1\int_0^\infty\left(1-\frac{2x_2}{3}+\frac{2x_2^2}{27}\right)
e^{-x_2\left(2/3-\frac{2m_e\eta}{W}\right)} g_{3S}(x_1,x_2)dx_2,
\end{displaymath} 
\begin{equation}
\label{vpvp1s2s}
\Delta E^{vp,vp}_{sopt}(3S-1S)=1.9955~meV.
\end{equation}

When calculating other corrections in the second order PT with the Breit potential, 
transformations of the original matrix elements are used to bring them to a form convenient 
for the integration. So, for example, when calculating the contribution of the amplitude 
in Fig.~\ref{fig1}(b), the following matrix element (n is the principal quantum number) 
appears:
\begin{equation}
\label{soptrel1}
M_{nS}=<\psi_{nS}|\frac{{\bf p}^4}{(2\mu)^2}{\sum}'_m\frac{|\psi_m><\psi_m|}{E_n-E_m}
\Delta V^C_{vp}|\psi_{nS}>=<\psi_{nS}|(E_n+\frac{Z\alpha}{r})(\hat H_0+
\frac{Z\alpha}{r})\times
\end{equation}
\begin{displaymath}
{\sum}'_m\frac{|\psi_m><\psi_m|}{E_n-E_m}\Delta V_{vp}^C |\psi_{nS}>
=<\psi_{nS}|\left(E_n+\frac{Z\alpha}{r}\right)^2\tilde G\Delta V_{vp}^C|\psi_{nS}>-
\end{displaymath}
\begin{displaymath}
<\psi_{nS}|\frac{Z\alpha}{r}\Delta V_{vp}^C |\psi_{nS}>+<\psi_{nS}|\frac{Z\alpha}{r}|\psi_{nS}>
<\psi_{nS}|\Delta V_{vp}^C|\psi_{nS}>.
\end{displaymath}
After integration over coordinates, an integral expression of the form is obtained:
\begin{equation}
\label{soptrel11}
M_{1S}=\frac{\mu^2\alpha(Z\alpha)^4}{12\pi}\int_1^\infty\frac{\rho(s)ds}{({b_1s}+1)^3}
\Biggl[
13 + 25 b_1 s + 8 b_1^2 s^2 + 8 (1 + b_1 s) \ln(1+b_1 s)
\Biggr],
\end{equation}
\begin{equation}
\label{soptrel22}
M_{3S}=\frac{\mu^2\alpha(Z\alpha)^4}{972\pi}\int_1^\infty\frac{\rho(s)ds}{ (3{b_1} s+1)^7}
\Biggl[52488 b_1^6 s^6+50301 b_1^5 s^5+23571 b_1^4 s^4+21546 b_1^3 s^3+
\end{equation}
\begin{displaymath}
1998 b_1^2 s^2+72 \left(729 b_1^5
s^5+243 b_1^4 s^4+162 b_1^3 s^3+54 b_1^2 s^2+3 b_1 s+1\right)\ln(3 b_1 s+1)+651 b_1 s+37 \Biggr].
\end{displaymath}

Another term in the Breit potential, proportional to $\delta({\bf r}) $, 
gives the Green's function with one zero argument $ \tilde G_{nS} ({\bf r},0) $ 
when calculating the matrix elements. The structure of the resulting expression 
after coordinate integration is quite similar to \eqref{soptrel11}-\eqref{soptrel22}:
\begin{equation}
\label{delta1}
\Delta E^{B,vp}_2(1S)=\frac{\mu^3\alpha(Z\alpha)^4}{6\pi}
\left(\frac{1}{m_1^2}+\frac{1}{m_2^2}\right)
\int_1^\infty\frac{\rho(s)ds}{(1+{b_1s})^3}
\Biggl[2{b_1}^2s^2+7{b_1}s+2({b_1}s+1) \ln({b_1}s+1)+3\Biggr],
\end{equation}
\begin{equation}
\label{delta2}
\Delta E^{B,vp}_2(3S)=\frac{\mu^2\alpha(Z\alpha)^4}{3\pi}
\left(\frac{1}{m_1^2}+\frac{1}{m_2^2}\right)
\int_1^\infty\frac{\rho(s)ds}{(1+2{b_1} s)^7}
\Biggl[-2 (3 b_1s+1) \left(243 b_1^4s^4+54 b_1^2s^2+1\right)\times
\end{equation}
\begin{displaymath}
\ln (3 b_1+1)-3 b_1s 
(6 b_1s (b_1s (3 b_1s(9 b_1s(3b_1s+2)+4)+28)+1)+5)-1\Biggr].
\end{displaymath}
Finally, the third term from \eqref{breit} gives, in the second order, 
the correction for recoil, which we represent in integral form as 
\eqref{soptrel11}, \eqref{soptrel22}, \eqref{delta1}, \eqref{delta2}:
\begin{equation}
\label{breit31s}
\Delta E^{B,vp}_3(1S)=-\frac{\mu^3 \alpha (Z\alpha)^4}{3\pi m_1m_2}\int_1^\infty
\frac{\rho(s)ds} {(1+{b_1} s)^3}
\Bigl[5{b_1} s+4 ({b_1} s+1) \ln ({b_1} s+1)+3\Bigr],
\end{equation}
\begin{equation}
\label{breit32s}
\Delta E^{B,vp}_3(3S)=-\frac{\mu^3 \alpha (Z\alpha)^4}{3\pi m_1m_2}\int_1^\infty
\frac{\rho(s)ds} {(1+2{b_1} s)^7}
\Bigl[1701 b_1^5s^5-567 b_1^4s^4+1998 b_1^3s^3+54 b_1^2s^2+
\end{equation}
\begin{displaymath}
12 \left(729 b_1^5s^5+243 b_1^4s^4+162 b_1^3s^3+54 b_1^2s^2+3
b_1s+1\right) \ln(3 b_1s+1)+69 b_1s+5\Bigr].
\end{displaymath}

\begin{figure}[t!]
\centering
\includegraphics[scale=0.8]{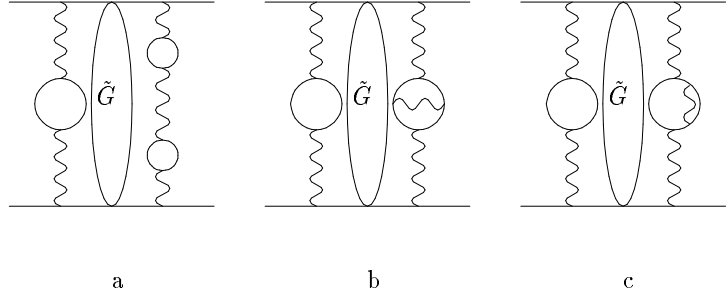}
\caption{Three-loop vacuum polarization corrections in the second order PT.
$\tilde G$ is the reduced Coulomb Green's function.}
\label{fig5}
\end{figure}

Numerical values of the contributions in Fig.~\ref{fig1} are shown in several 
lines 10, 11, 12. Since the contribution of the interaction in Fig.~\ref{fig1}(c) 
has the order $\alpha^2(Z\alpha)^2$, then the addition of one VP loop leaves 
such a correction potentially important. The contribution of the three-loop VP in the 
second order PT is shown in Fig.~\ref{fig5} (all perturbation potentials 
are the corrections of the vacuum polarization to the Coulomb potential). Omitting the details 
of the calculation of this contribution (see \cite{apm2015}), since they are similar 
to the calculation of the amplitude in Fig.~\ref{fig1} (c), we present 
its numerical value in the Table~\ref{tb1} (line 13).
 
The contribution of the three-loop vacuum polarization in the third order PT
is shown in Fig.~\ref{fig33_1} (c). Analytically, it is determined 
by the sum of two terms \cite{ikk2009}:
\begin{equation}
\label{eq:43a}
\Delta E_{nS}=<\psi_{nS}|\Delta V^C\tilde G\Delta V^C\tilde G\Delta V^C|\psi_{nS}>-
<\psi_{nS}|\Delta V^C|\psi_{nS}><\psi_{nS}|\Delta V^C\tilde G\tilde G\Delta V^C|\psi_{nS}>.
\end{equation}
Corrections of this type are presented in the form of multiple integrals over spectral 
parameters as in \cite{apm2019} and calculated numerically (see line 14 
of Table~\ref{tb1}). Numerical integration is performed with good accuracy, accepted in the work.

\section{Corrections to the nuclear structure and vacuum polarization}

\begin{figure}[htbp]
\centering
\includegraphics[scale=0.8]{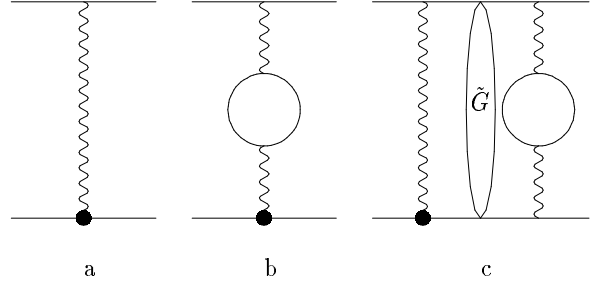}
\caption{Corrections to the nuclear structure and the polarization of vacuum 
in the first order PT (FOPT) and the second order PT (SOPT).}
\label{fig6}
\end{figure}

\begin{figure}[t!]
\centering
\includegraphics[scale=0.8]{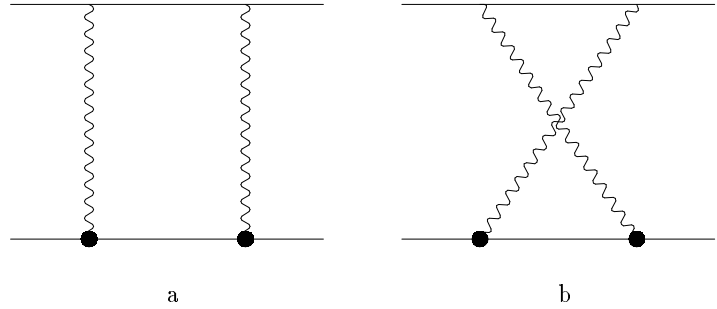}
\caption{Corrections to the nuclear structure of order $ (Z\alpha)^5 $. 
The bold point in the diagram denotes the vertex operator of the proton.}
\label{fig7}
\end{figure}

A decrease in the value of the Bohr radius of orbits in muonic atoms in comparison with 
electronic ones leads to the fact that the wave function of the muon overlaps strongly 
with the region of the proton. Expanding the charge form factor of the proton at small 
momentum transfers, we find that in the leading order the effect of the nuclear structure 
is determined in the energy range $(3S-1S)$ by the following correction proportional 
to the square of the charge radius $r_{p}^2$ \cite{egs} 
(see Fig.~\ref{fig6} (a)) (the subscript $str$ denotes here and below a correction 
for the nuclear structure, hereinafter, for the proton charge radius, we use the notation
$r_p$):
\begin{equation} 
\label{rn2}
\Delta E_{str}(3S-1S)=-\frac{52\mu^3(Z\alpha)^4}{81} r^2_{p} =
-40.039631~ r_{p}^2=-28.3105~meV,
\end{equation}
where we have extracted the coefficient at $r_{p}^2 $, and the value of the charge 
radius itself is taken in fm. For a numerical estimate of the contributions \eqref{rn2}, 
the value of the proton charge radius from \cite{crema} is used. The next most 
important correction for muonic hydrogen is the correction for the structure 
of the nucleus of order $(Z\alpha)^5$ of two-photon exchange amplitudes (see Fig.~\ref{fig7}), 
which is expressed in terms of the Dirac $F_1$ and Pauli $F_2$ form factors 
of the proton. Neglecting the relative momenta of particles in the initial 
and final states, one can represent this contribution to the shift of $S$ levels 
in the integral form:
\begin{equation}
\label{2gamma}
\Delta E_{str}^{2\gamma}(3S-1S)=\frac{26\mu^3(Z\alpha)^5}{27\pi}\delta_{l0}
\int_0^\infty\frac{dk}{k}V_{2\gamma}(k),~
V_{2\gamma}(k)=\frac{2(F_1^2-1)}{m_1m_2}+\frac{8m_1[F_2(0)+4m_2^2F_1'(0)]}{m_2(m_1+m_2)k}+
\end{equation}
\begin{eqnarray}
+\frac{k^2}{2m_1^3m_2^3}\bigl[2(F_1^2-1)(m_1^2+m_2^2)+4F_1F_2m_1^2+3F_2^2m_1^2\bigr]+
\frac{\sqrt{k^2+4m_1^2}}{2m_1^3m_2(m_1^2-m_2^2)k}\times \nonumber \\
\left\{k^2[2(F_1^2-1)m_2^2+4F_1F_2m_1^2+3F_2^2m_1^2]-8m_1^4F_1F_2+
\frac{16m_1^4m_2^2(F_1^2-1)}{k^2}\right\}-\nonumber  \\
-\frac{\sqrt{k^2+4m_2^2}m_1}{2m_2^3(m_1^2-m_2^2)k}\left\{k^2[2(F_1^2-1)+4F_1F_2+3F_2^2]-
8m_2^2F_1F_2+\frac{16m_2^4(F_1^2-1)}{k^2}\right\},  \nonumber
\end{eqnarray}
In numerical integration in \eqref{2gamma}, we use the parametrization of the proton 
form factors from \cite{kelly}.
Numerical result of the correction for the nuclear structure from two-photon exchange 
amplitudes \eqref{2gamma} is included in Table~\ref{tb1} in line 18. The error 
in calculating this contribution is estimated at $1\% $, therefore the result is given 
with an accuracy of two digits after the decimal point. The magnitude of this correction 
in muonic hydrogen increases significantly in comparison with electron hydrogen. 
The contribution of two-photon exchange amplitudes in the case of a point proton 
is known from the calculation in muonium and is presented 
in the next section.

Contributions of the 5th order in $\alpha$ also give the amplitudes of particle 
interactions, which contain both the effects of the nuclear structure and vacuum 
polarization (see Fig.~\ref{fig6}). The particle interaction operator 
in the coordinate representation corresponding to the diagram in 
Fig.~\ref{fig6}(b) has the form:
\begin{equation}
\label{eq8a}
\Delta V^{vp}_{str}(r)=\frac{2}{3}\pi Z\alpha r^2_p \frac{\alpha}{3\pi}
\int_1^\infty\rho(\xi)d\xi\left[\delta({\bf r})-\frac{m_e^2\xi^2}{\pi r}
e^{-2m_e\xi r}\right].
\end{equation}
Using the expression \eqref{eq8a}, you can perform analytical integration 
over all variables when calculating the matrix elements. For $1S$ and $3S$ states, 
the energy level shifts are equal:
\begin{equation}
\label{vpstr1s}
\Delta E^{vp}_{str}(1S)=\frac{2\alpha(Z\alpha)^4r_{p}^2\mu^3}{27\pi\sqrt{1-b_1^2}}
\Bigl[
\left(6 b_1^4-3 b_1^2+6\right) \ln \left(\frac{\sqrt{1-b_1^2}+1}
{b_1}\right)+\sqrt{1-b_1^2}\left(-3 \pi  b_1^3+6 b_1^2+1\right)
\Bigr],
\end{equation}
\begin{equation}
\label{vpstr2s}
\Delta E^{vp}_{str}(3S)=\frac{\alpha(Z\alpha)^4r_p^2\mu^3}{1458\pi 
\bigl(1-9 b_1^2\bigr)^{9/2}}\Bigl\{\sqrt{1-9 b_1^2}\bigl[
32+9 b_1^2(-116+9 b_1(3b_1(19+6b_1^2(-5-414 b_1^2+972 b_1^4))-
\end{equation}
\begin{displaymath}
4 (1-9 b_1^2)^4 \pi))\bigr]+3(8+81 b_1^2(-4+9 b_1^2(10+b_1^2(-89+162 b_1^2
(5-18b_1^2+36 b_1^4)))))\ln\frac{3b_1}{1-\sqrt{1-9 b_1^2}}
\Bigr\}.
\end{displaymath}

The contribution to the energy spectrum of the same order $\alpha (Z\alpha)^4 $ 
is determined by the same effects in the second order of the perturbation theory 
(see Fig.~\ref{fig6}(c)) by the following integral expressions:
\begin{equation}
\label{vpstr1}
\Delta E^{vp}_{str,sopt}(1S)=\frac{2\alpha(Z\alpha)^4\mu^3 r_p^2}
{9\pi}\int_1^\infty\frac{\rho(\xi)d\xi}{ ({b_1} \xi+1)^3}
\bigl[2{b_1}^2 \xi^2-7{b_1}\xi-2({b_1}\xi+1)\ln({b_1}\xi+1)-3\bigr],
\end{equation}
\begin{equation}
\label{vpstr2}
\Delta E^{vp}_{str,sopt}(3S)=-\frac{4\alpha(Z\alpha)^4\mu^3 r_p^2}
{81\pi}\int_1^\infty\frac{\rho(\xi)d\xi}{(2{b_1} \xi+1)^7}
\Bigl[ \ln(1+3 b_1\xi)\times
\end{equation}
\begin{displaymath}
2(1+3 b_1\xi) (1+54 b_1^2\xi^2+243 b_1^4\xi^4)+
(1+3 b_1\xi (5+6 b_1\xi (1+b_1\xi (28+3b_1\xi (4+9b_1\xi (2+3 b_1\xi))))))
\Bigr].
\end{displaymath}

\begin{figure}[t!]
\centering
\includegraphics[scale=0.8]{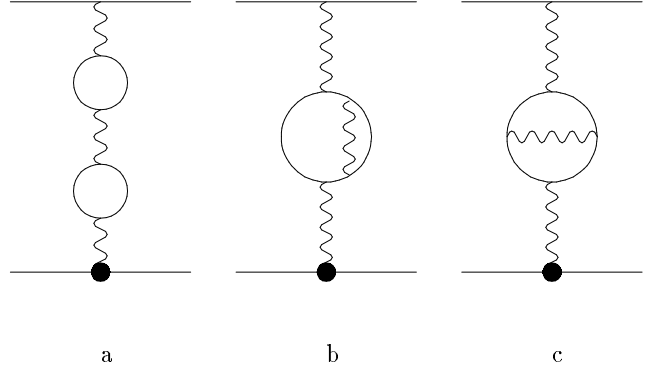}
\caption{Nuclear structure effects and two-loop vacuum polarization in one-photon interaction. 
The bold point in the diagram denotes the proton vertex operator.}
\label{fig8}
\end{figure}

In Table~\ref{tb1}, line 16 includes the total contribution of \eqref{vpstr1s} 
and \eqref{vpstr1}, \eqref{vpstr2s} and \eqref{vpstr2}. Since 
\eqref{vpstr1s}-\eqref{vpstr2} are multiplied by $ r_p^2 $ and are numerically 
large, they can be combined with \eqref{rn2} to increase the accuracy of extracting 
the proton charge radius in the presence of experimental data. The corrections 
for the two-loop vacuum polarization taking into account the structure of the 
proton are calculated in the same sequence. In one-photon interaction, they 
are shown in Fig.~\ref{fig8}(a,b,c). The particle interaction operators 
corresponding to these amplitudes are constructed in the integral form:
\begin{equation}
\label{vp-vp-str}
\Delta V^{vp-vp}_{str}(r)=\frac{2}{3}Z\alpha r_p^2 \left(\frac{\alpha}
{3\pi}\right)^2\int_1^\infty\rho(\xi)d\xi\int_1^\infty\rho(\eta)d\eta\times
\end{equation}
\begin{displaymath}
\times\left[\pi\delta({\bf r})-\frac{m_e^2}{r(\xi^2-\eta^2)}\left
(\xi^4 e^{-2m_e\xi r}-\eta^4e^{-2m_e\eta r}\right)\right],
\end{displaymath}
\begin{equation}
\label{2loopvp-str}
\Delta V^{2-loop~vp}_{str}(r)=\frac{4}{9}Z\alpha r_p^2 \left
(\frac{\alpha}{\pi}\right)^2\int_0^1\frac{f(v)dv}{1-v^2}\left[\pi\delta({\bf r})-
\frac{m_e^2}{r(1-v^2)}e^{-\frac{2m_er}{\sqrt{1-v^2}}}\right].
\end{equation}

\begin{figure}[t!]
\centering
\includegraphics[scale=0.8]{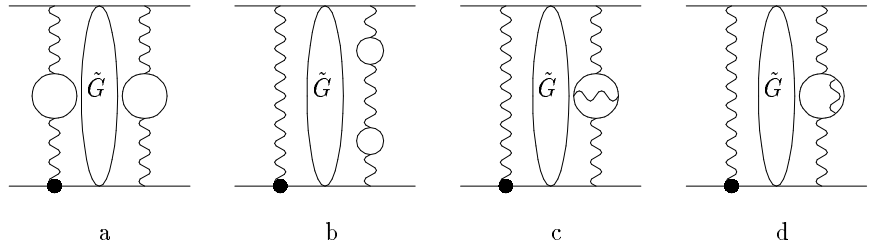}
\caption{Nuclear structure and two-loop vacuum polarization effects
in the second order of perturbation theory. The bold point in the diagram denotes 
the proton vertex operator. $ \tilde G $ is the reduced Coulomb Green's function.}
\label{fig9}
\end{figure}

In the second order PT, the contributions of the two-loop vacuum polarization with 
the effect of the nuclear structure
of order $\alpha^2(Z\alpha)^4$ are shown in Fig.~\ref{fig9}(a,b,c,d). 
When calculating the contribution in Fig.~\ref{fig9}(a), it is convenient 
to split it into two parts when integrating over the coordinates of particles 
in accordance with the two terms of the potential \eqref{eq8a}. Each of them 
diverges upon integration over spectral parameters, but their sum is finite. 
The other corrections in Fig.~\ref{fig9}(b,c,d) are calculated in the same way, 
and their sum is presented in Table~\ref{tb1} (line 17). Integral expressions 
for these corrections before integration over spectral parameters are rather 
cumbersome and are not presented here (see, for example, \cite{apm2019}).
\begin{figure}[t!]
\centering
\includegraphics[scale=1.]{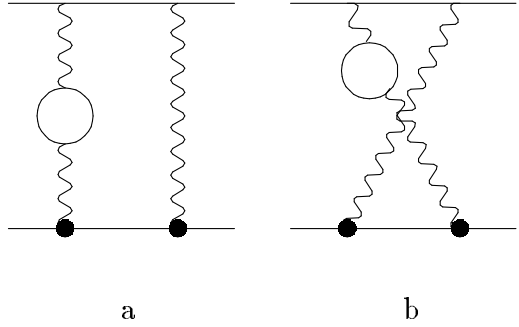}
\caption{Nuclear structure and vacuum polarization effects in two-photon exchange 
diagrams. The bold point in the diagram denotes the proton vertex operator.}
\label{fig10}
\end{figure}

\begin{figure}[t!]
\centering
\includegraphics[scale=0.8]{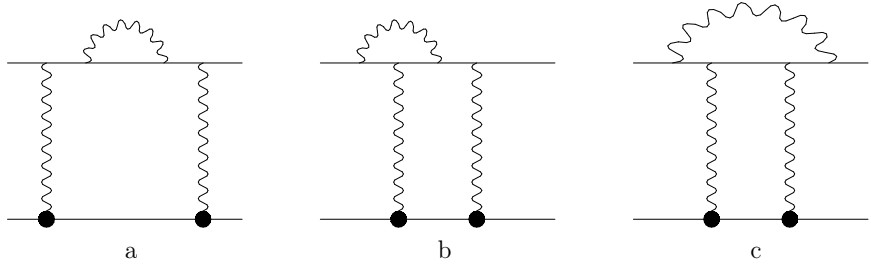}
\caption{Two-photon exchange amplitudes with radiative corrections to the muon line, 
contributing to the order $\alpha(Z\alpha)^5$. The bold point 
in the diagram shows the proton vertex operator.}
\label{fig12}
\end{figure}

We also take into account in our calculation the combined sixth-order $\alpha$ 
correction for the nuclear structure and vacuum polarization, which appears in 
two-photon exchange amplitudes as a result of the modification of the photon 
propagator (see Fig.~\ref{fig10}). The corresponding particle interaction 
operator differs from $ V_{2\gamma}(k) $ from
\eqref{2gamma} with an additional functional factor. The integral expression 
for the correction of this type is \cite{apm2005,apm2007,apm2018}:
\begin{equation}
\label{2gammavp}
\Delta E^{2\gamma}_{str,vp}(nS)=-\frac{2\mu^3\alpha(Z\alpha)^5}{9\pi^2 n^3} \int_0^\infty 
\frac{V_{2\gamma}(k) F_{vp}(k)dk}{k^4},
\end{equation}
\begin{displaymath}
F_{vp}(k)=\Bigr[-5 k^3+6 \left(k^2-2 {m_e}^2\right) \sqrt{k^2+4{m_e}^2} \tanh^{-1}\left(\frac{k}
{\sqrt{k^2+4{m_e}^2}}\right)+12 k{m_e}^2\Bigl],
\end{displaymath}
and its numerical value for the $(3S-1S)$ interval is given in Table~\ref{tb1} (line 19). 
Other radiative corrections to the muon line (self-energy correction (se), vertex correction 
and correction with an enveloping photon) with a nuclear structure of the same order 
$\alpha(Z\alpha)^5$ are determined by the amplitudes in Fig.~\ref{fig12}. 
Their calculation is performed for $S$-states in light muonic atoms in \cite{plb2017}. 
The total contribution, expressed in terms of the nuclear charge radius, is:
\begin{equation}
\label{2gamma_rad_str}
\Delta E^{\alpha(Z\alpha)^5}_{str,rad}(3S-1S)=\frac{13\mu^3\alpha(Z\alpha)^5 r_p^2}{81} 
(23-16\ln 2),
\end{equation}
and numerical values are shown in line 20 of the Table~\ref{tb1}.

\section{Recoil corrections, muon self-energy and vacuum polarization corrections}

So far, we have considered corrections of various orders in the $(3S-1S)$ interval, 
which are specific for each muonic atom. Such contributions were obtained either 
analytically or in the form of integral expressions and calculated numerically. 
In muonic atoms, expansion in terms of the characteristic parameter 
$\mu Z\alpha/m_e=1.356$ cannot be used. For the electron hydrogen atom, 
you can also use these expressions, replacing $ m_e\to m_\mu $. But there 
is also another set of contributions that are known in analytical form and were obtained 
in the study of the hydrogen atom fine structure energy spectrum \cite{egs}. They can be used 
to estimate numerically the contributions in the case of muonic atoms. 
Let us briefly list the main contributions required to obtain a complete result with good accuracy
\cite{egs,kp1996,friar,borie,apm2019}.

There is a group of recoil corrections (the subscript $rec$ is used) of various orders in 
$(Z\alpha)$, obtained in the case of a point nucleus. The recoil contribution 
of order $(Z\alpha)^4m_1^2/m_2^2 $ arises when calculating 
the matrix elements of the Breit potential. For the hydrogen atom, such corrections 
are taken into account in the original formula \eqref{eq1}. The recoil correction 
of order $(Z \alpha)^5m_1/ m_2 $ for $S$-states is determined 
by two-photon exchange amplitudes, in which the proton is considered as a point particle 
\cite{egs}:
\begin{equation}
\label{eq58}
\Delta E_{rec}^{(Z\alpha)^5}=\frac{\mu^3(Z\alpha)^5}{m_1m_2\pi n^3}\Bigl[\frac{2}{3}
\ln\frac{1}{Z\alpha}-\frac{8}{3}\ln k_0(n,0)-\frac{1}{9}-\frac{7}{3}a_n-
\frac{2}{m_2^2-m_1^2}(m_2^2\ln\frac{m_1}{\mu}-m_1^2\ln\frac{m_2}{\mu})\Bigr],
\end{equation}
where $\ln k_0(n,0)$ is the Bethe logarithm, which has the following values for $1S$, $3S$ states
\cite{egs}:
\begin{equation}
\label{eq59}
\ln k_0(1S)=2.984128555765498,
~~~\ln k_0(3S)=2.767663612491822,
\end{equation}
\begin{equation}
\label{eq:65}
a_n=-2\left[\ln\frac{2}{n}+(1+\frac{1}{2}+...+\frac{1}{n})+1-\frac{1}{2n}\right].
\end{equation}
The numerical value \eqref{eq58} for the interval $(3S-1S)$ is large for both muonic 
and electron hydrogen (line 22, Table~\ref{tb1}).

The recoil correction of order $(Z\alpha)^6m_1/m_2 $ has been studied 
in many works \cite{eides1, pg, shabaev}, and in \cite{erokhin2015} 
the corrections for the recoil of a higher order $m_1^2 (Z\alpha)^7/m_2 $ are calculated. 
Since in our work we limited ourselves to contributions to the sixth order in $Z\alpha $, 
we use the following expression to obtain a numerical estimate (see \cite{egs}):
\begin{equation}
\label{eq61}
\Delta E_{rec}^{(Z\alpha)^6}(3S-1S)=\frac{26(Z\alpha)^6m_1^2}{27m_2}
\left(\frac{7}{2}-4\ln 2\right).
\end{equation}
In the case of muonic hydrogen, the numerical value of the correction of order
$m_1^2(Z\alpha)^7/m_2$ is negligible \cite{erokhin2015}.

For the energy contributions obtained from amplitudes with radiative corrections 
to the muon line (se), from the Dirac and Pauli form factors (ff) of the muon, there 
is a compact analytical representation \cite{egs}:
\begin{equation}
\label{eq62}
\Delta E_{se,ff}(nS)=\frac{\alpha(Z\alpha)^4}{\pi n^3}\frac{\mu^3}{m_1^2}
\Biggl(\frac{4}{3}\ln\frac{m_1}{\mu(Z\alpha)^2}-\frac{4}{3}\ln k_0(n,0)+
\frac{10}{9}+
\end{equation}
\begin{displaymath}
+\frac{\alpha}{\pi}\left(-\frac{9}{4}\zeta(3)+\frac{3}{2}
\pi^2\ln 2-\frac{10}{27}\pi^2-\frac{2179}{648}\right)+4\pi Z\alpha\left(
\frac{427}{384}-\frac{\ln 2}{2}\right)\Biggr).
\end{displaymath}
A discussion of higher-order contributions in $\alpha$ can be found in 
\cite{shabaev2015} (see also references to other articles in this paper).

Radiative corrections with recoil of orders $\alpha (Z\alpha)^5 $ and 
$(Z^2\alpha) (Z\alpha)^4$ from Table~9 \cite{egs} have the following form for $nS$ states:
\begin{equation}
\label{eq63}
\Delta E_{rad-rec}(nS)=
\frac{\alpha(Z\alpha)^5\mu^3}{m_1m_2n^3}
(6\zeta(3) - 2\pi^2\ln 2 + \frac{3}{4}\pi^2 - 14)+
\frac{\alpha(Z\alpha)^5m_1^2}{m_2n^3\pi^2}\left(\frac{2\pi^2}{9}-\frac{70}{27}\right)+
\end{equation}
\begin{displaymath}
+\frac{4(Z^2\alpha)(Z\alpha)^4\mu^3}{\pi m_2^2n^3}
\left(\frac{1}{3}\ln\frac{\Lambda(Z\alpha)^{-2}}{\mu}+\frac{11}{72}-\frac{1}{24}-\frac{7\pi}{32}
\frac{\Lambda^2}{4m_2^2}+\frac{2}{3}\left(\frac{\Lambda^2}{4m_2^2}\right)^2-\frac{1}{3}
\ln k_0(n,0)\right).
\end{displaymath}
The formulas \eqref{eq62} and \eqref{eq63} contribute to the shift $(3S-1S)$ 
(parameter $\Lambda=\sqrt{12/r_p^2}$), which is shown in Table~\ref{tb1} 
on separate lines 24-25.

The correction for the proton structure in the energy spectrum of order $(Z\alpha)^6$, obtained 
in \cite{friar} for a muonic hydrogen-like atom with various parametrizations 
of the nuclear form factors, has the form:
\begin{equation}
\label{eq64}
\Delta E_{str}^{(Z\alpha)^6}(nS)=\frac{2\mu(Z\alpha)^6}{3n^3}\left[\mu^2 F_{rel}+\mu^4 F_{nr}\right],
\end{equation}
\begin{equation}
\label{frel}
F_{rel}=-<r^2>\Bigl[\psi(n)+2\gamma+\frac{9}{4n^2}-\frac{1}{n}-\frac{13}{4}+<\ln\frac{2Wr}{n}>\Bigr]
-\frac{1}{3}<r^3><\frac{1}{r}>+I_2^{rel}+I_3^{rel},
\end{equation}
\begin{equation}
\label{fnr}
F_{nr}=\frac{2<r^2>}{3}\Bigl[<r^2>\left(
\psi(n)+2\gamma-\frac{1}{n}-\frac{4}{3}\right)+<r^2\ln\frac{2Wr}{n}>\Bigr]+
\end{equation}
\begin{displaymath}
\frac{<r^4>}{10n^2}+
<r^3><r>+<r^5><\frac{1}{r}>+I_2^{nr}+I_3^{nr},
\end{displaymath}
where the quantities $I_{2,3}^{rel}$ and $I_{2,3}^{nr} $, as well as the moments 
of the charge distribution density, are written explicitly in \cite{friar} 
for various parameterizations. Based on the expressions obtained in \cite{friar}, 
one can estimate the contributions to the shift $(3S-1S)$ for muonic hydrogen atom 
(line 26 of the Table).

\begin{figure}[t!]
\centering
\includegraphics{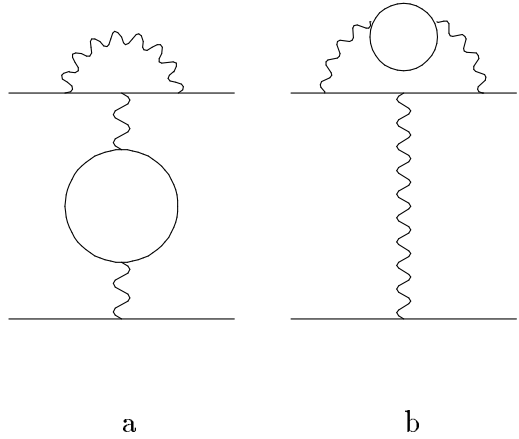}
\caption{Radiative corrections with vacuum polarization effects.}
\label{fig13}
\end{figure}

Another numerically important contribution of the sixth order in $\alpha $ 
to the shift $(3S-1S)$ (see Fig.~\ref{fig13} (b)) is expressed in terms 
of the slope of the Dirac form factor $F_1'(0)$ and Pauli form factor $F_2(0)$ 
\cite{egs}, which are calculated analytically in \cite{bcr} (see line 27 of Table~\ref{tb1}):
\begin{equation}
\label{eq67}
\Delta E_{rad+vp}=-\frac{7\alpha^2(Z\alpha)^4\mu^3}{8\pi^2m_1^2}\Bigl[
\frac{3{m_e}^2}{m_1^2}-\frac{4{m_e}^2 \ln\frac{m_1 }{{m_e}}}{m_1^2}+
\frac{\pi^2{m_e}}{4m_1 }+\frac{4}{9}
\ln^2\frac{m_1}{{m_e}}-\frac{20}{27} \ln\frac{m_1}{{m_e}}+
\frac{2\pi^2}{27}+\frac{85}{162}\Bigr].
\end{equation}

In the case of electron hydrogen, a numerically large contribution is giving 
by radiative corrections of order $\alpha(Z\alpha)^6$ (see Table~5 from \cite{egs}). 
We have included it in our Table~\ref{tb1} for muonic hydrogen in line 31.

To estimate the contribution of the muon self-energy (mse) taking into account 
the vacuum polarization, the following expression was obtained 
in the logarithmic approximation in \cite{kp1996}:
\begin{equation}
\label{eq68}
\Delta E^{vp}_{mse}(n)=\frac{\alpha}{3\pi
m_1^2}\ln\frac{m_1}{\mu(Z\alpha)^2} \left[<\psi_n|\Delta\cdot 
\Delta V^C_{vp}|\psi_n>+2<\psi_n|\Delta V^C_{vp}\tilde G
\Delta\left(-\frac{Z\alpha}{r}\right)|\psi_n>\right].
\end{equation}
Assuming as in calculating relativistic corrections ${\bf p}^2=2 \mu (H + Z\alpha/r)$ 
and calculating numerous matrix elements in \eqref{eq68}, we obtain a correction 
in the interval $(3S-1S)$ (line 28).

Taking into account the accuracy of the calculation, we have included in the complete 
result for the shift $(3S-1S)$ the contribution of hadronic vacuum polarization (hvp), 
which was investigated in \cite{borie1981,friar2,m6} in the case of muonic hydrogen. 

\begin{figure}[t!]
\centering
\includegraphics[scale=0.8]{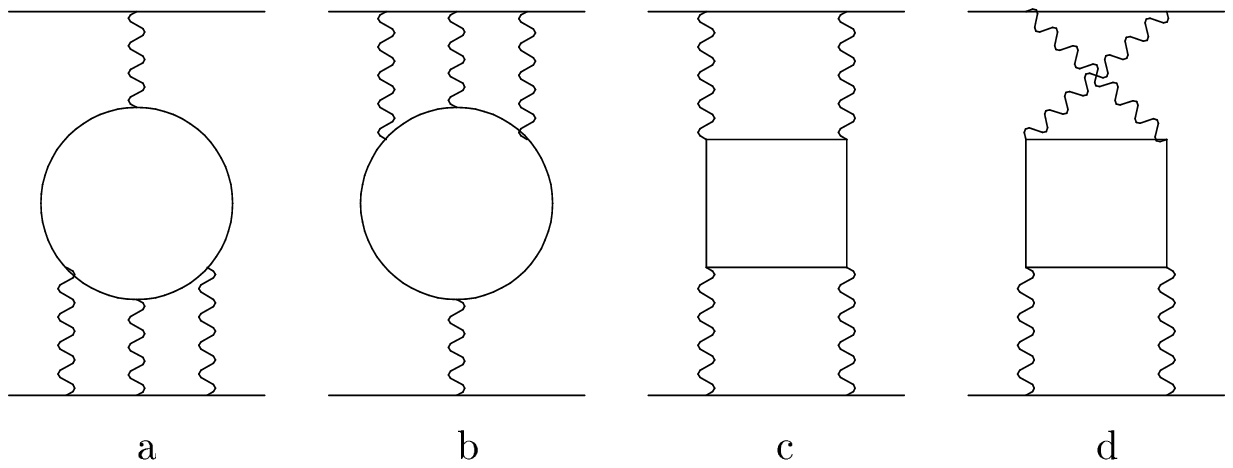}
\caption{The amplitudes of light-by-light scattering.}
\label{figll}
\end{figure}

Fig.~\ref{figll} shows four amplitudes of light-by-light scattering. The amplitude 
in Fig.~\ref{figll}(a) denotes the contribution known as the Wichmann-Kroll correction 
(see the approximation potential in \cite{egs}). It is shown in Table~\ref{tb1} in a separate 
line 4. In the works \cite{sgk2018, korzinin} it was shown that the contribution 
of the amplitude in Fig.~\ref{figll} differs from the Wichmann-Kroll contribution 
by the factor $ 1/Z^2$. The interaction potential in Fig.~\ref{figll}(c, d) 
was obtained in \cite{sgk2018} using the Pade approximation in a convenient form 
for numerical calculations of corrections in the energy spectrum (the coefficients 
$s_i$, $t_i$ are written in \cite{sgk2018}) (index $ll$ corresponds 
to the abbreviation light-by-light):
\begin{equation}
\label{ll}
\Delta V^{ll}(r)=-\frac{\alpha^2(Z\alpha)^2}{r}\frac{(s_0+s_1x+s_2x^2)}{(t_0+t_1x+t_2x^2+t_3x^3+
t_4x^4+t_5x^5)},~~~
x=m_e r.
\end{equation}
Numerical value of the corresponding correction (see line 5, in which the total 
contribution of the amplitudes in Fig.~\ref{figll}(b,c,d) is written out) 
is important for refining the complete result.

There is one more effect of light-by-light scattering, which leads to the appearance 
of an effective one-meson interaction between a muon and a proton (scalar, pseudoscalar, 
axial-vector and tensor). Studies of such 
a mechanism in \cite{aed,aed1,aed2} have shown that in muonic hydrogen, the exchange 
of a scalar meson gives a significant shift of the $S$-energy levels 
$\Delta E(3S-1S)=-0.1059~meV$. Therefore, it was also included in the final 
Table~\ref{tb1} for muonic hydrogen.

\section{Numerical results and conclusion}
\label{sec:concl}

In this work, we continue our studies \cite{apm2005,apm2015,apm2019,apm2007} of low-lying
energy levels of muonic hydrogen, which have been intensively studied in recent years experimentally 
and theoretically. The $(3S-1S)$ transition was chosen as the energy interval for precision calculation, 
the measurement accuracy of which in the case of electron hydrogen is very high 
\cite{fleurbaey}. Various potentially important interactions in muonic hydrogen were analyzed 
and their contribution to the structure of $S$-states was calculated within the framework 
of the quasipotential method in quantum electrodynamics. We have calculated the energy interval 
$(3S-1S)$ in muonic hydrogen taking into account corrections of the fifth and sixth orders 
in $\alpha$ and taking into account the effects of recoil and the structure of the proton. 
We also used numerous results of calculations of other corrections performed by different authors 
and took into account numerically important effects of higher order in $\alpha$.

The interest in the transitions between the $S$-energy levels in the hydrogen atom is due 
to the fact that the study of such transitions opens up yet another possibility 
of refining the value of the charge radius of the nucleus (proton). As noted earlier, 
there is the experiment \cite{fleurbaey} that is quite consistent with the CODATA result 
for the proton charge radius, unlike all other new experiments with both muonic and 
electronic systems. In this regard, it was useful to calculate the energy interval 
$(3S-1S)$ in muonic hydrogen, bearing in mind the possible perspective of its measurement 
in the future.

In recent years, different groups have performed calculations of new corrections 
in the fine and hyperfine structure of the spectrum of muonic atoms, including for 
the $S$-energy levels. Since the number of such papers is significant, references 
to many other papers can be found in the review papers \cite{crema2016,crema,egs,borie,franke}. 
Although these works did not directly calculate the energy interval $(3S-1S)$, some comparison 
with the results obtained earlier can be made. Most of the corrections we calculated for vacuum 
polarization (Uhling, K\"allen-Sabri, Wichmann-Kroll corrections) are in good agreement 
with previous calculations \cite{crema,crema2016,borie,sgk,sgk1}, but in this work these corrections 
are not presented in summary form, but in more detail. Our calculations 
of three-loop VP effects in one-photon interaction are based on the Kinoshita-Nio 
interpolation formula (8 diagrams in Fig.~\ref{fig22}), which gives the known 
result for the Lamb shift from \cite{kn}. For the interval $(3S-1S)$, this contribution 
was obtained for the first time, as well as the contribution of the three-loop polarization 
operator with two fermionic cycles in Fig.~\ref{fig33}, calculated on the basis 
of the analytical formula for the polarization operator from \cite{hoang,hoang1}.
We have worked out in detail the calculation of combined corrections for the vacuum polarization 
with relativistic effects. We have written down the corresponding potentials and matrix elements 
in the first and second orders of the perturbation theory.
New expression is obtained for the reduced Coulomb Green's function of $3S$-state 
with two nonzero arguments, 
which is necessary for calculating corrections in the second and third orders 
of the perturbation theory. General expressions for calculating the effects of a three-loop 
VP in the third order of perturbation theory agree with \cite{sgk,sgk1}, and numerical 
results themselves are new, as well as numerous corrections for the proton structure. 
Note that to calculate the effects of light-by-light scattering, 
we use the formula for the potential obtained in \cite{sgk2018}. The calculation of corrections 
in Section~5 is based on well-known analytical expressions (see \cite{egs}) and the results 
are presented in detail in Table~\ref{tb1}.

The calculated contributions are presented in the work in the form of analytical formulas, 
integral expressions that can be integrated numerically, as well as in numerical form 
in the Table~\ref{tb1}. Most of the results for muonic hydrogen in the Table~\ref{tb1}
are presented with an accuracy of four digits after the decimal point in meV, since 
the errors in their determination due to the errors of fundamental physical constants 
are much smaller. But there are contributions to the proton structure, which 
are determined by the strong interaction and have been obtained so far with a significant error. 
Total theoretical result for the energy interval $(3S-1S)$ in muonic hydrogen has the form
$\Delta E(\mu p)=2249398.5478 $ meV.
The theoretical error in the calculation of the interval $(3S-1S)$ connected with the calculation 
of the QED corrections is, according to our estimates, about 0.005 meV. Our estimate 
of the contribution to the proton polarizability is taken from the work \cite{apm2006}. 
The largest contribution to the structure of the proton is connected with a correction 
of order $(Z\alpha)^4$ (line 15 of the Table~\ref{tb1}). 
Its numerical value is obtained with the proton charge radius from \cite{crema}. 
If we do not fix numerical value of the correction \eqref{rn2}, 
then the total result from the Table~\ref{tb1} can be presented as:
\begin{equation} 
\label{rezt}
\Delta E^{tot}(3S-1S)=
2249426.8578-40.039631~ r_p^2~meV.
\end{equation}
Thus, a precision measurement of the frequency of the $(3S-1S)$ transition 
in muonic hydrogen can give a new, more the exact value of the proton's charge radius. 
So, for example, measuring the $(3S-1S)$ shift in muonic hydrogen with a relative 
error $(1\div 3)$ ppb will reduce the error in determining the proton charge radius to 0.0001 fm.

\acknowledgements
This work was supported by the Russian Science Foundation (grant No. 18-12-00128).

\newpage 

\begin{table}[htbp]
\centering
\caption{Corrections to energy interval $(3S-1S)$ in muonic hydrogen}
\label{tb1}
\begin{ruledtabular}
\begin{tabular}{|c|c|c|}  \hline
No.& Contribution to the interval $(3S-1S)$                  &$(\mu~p)$, meV    \\   \hline
1  &Fine structure correction \eqref{eq1}        & 2247582.5823    \\   \hline
2  & Vacuum polarization correction in $1\gamma$          &1834.5535      \\
   &interaction of order $\alpha(Z\alpha)^2$                       &                   \\    \hline
3  & Muon VP correction in $1\gamma$                     &   0.1276              \\        
   & interaction of order $\alpha(Z\alpha)^4$       &                       \\    \hline
4  & The Wichman-Kroll correction                        &-0.0110          \\     \hline
5  & Light-by-light correction               & 0.0044                  \\     \hline
6  & Two-loop VP correction in $1\gamma$                &   12.8432    \\
   & interaction of order $\alpha^2(Z\alpha)^2$     &                  \\    \hline
7  & Three-loop VP correction in $1\gamma$               &  0.0246             \\
   & interaction of order $\alpha^3(Z\alpha)^2$     &                     \\    \hline
8  & Relativistic corrections with                      &  -0.1804                   \\
   & the account of one-loop VP in FOPT                    &                     \\    \hline
9  &  Relativistic corrections with                       &   -0.0010         \\
   & the account of two-loop VP in FOPT                     &                   \\    \hline
10 & Relativistic corrections with                      &  0.2900             \\
   & the account of one-loop VP in SOPT                  &                   \\    \hline
11 & Relativistic corrections with                         & -0.0012                 \\
   & the account of two-loop VP in SOPT                  &                \\    \hline
12 &Two-loop VP correction in                       &  1.9955            \\
   & second order PT of order $\alpha^2(Z\alpha)^2$  &                  \\    \hline
13 &Three-loop VP correction in                      &  0.0294        \\
   & second order PT of order $\alpha^3(Z\alpha)^2$  &                     \\    \hline
14 &Three-loop VP correction in                        &  0.0032                   \\
   & third order PT of order $\alpha^3(Z\alpha)^2$ &                  \\    \hline
15 & Nuclear structure correction of order $(Z\alpha)^4$    &-28.31 $\pm 0.04$   \\   \hline
16 & Correction to the nuclear structure with   &   -0.1705     \\
   & the account of vacuum polarization of order $\alpha(Z\alpha)^4$    &    \\     \hline
17 & Correction to the nuclear structure  with            &  -0.0017          \\     
   & two-loop VP of order $\alpha^2(Z\alpha)^4$     &      \\     \hline
18 & Nuclear structure correction from       &  0.18       \\
   &$2\gamma$ amplitudes of order $(Z\alpha)^5$       &                 \\     \hline
19 & Nuclear structure and VP correction in   & 0.0033       \\
   &2$\gamma$ interaction of order $\alpha(Z\alpha)^5$      &           \\    \hline
20 &  Radiative corrections in muon line           &   0.0045       \\
   &  with nuclear structure of order $\alpha(Z\alpha)^5$   &        \\    \hline
21 & Proton polarizability correction     & $0.10\pm 0.02$   \\ \hline
\end{tabular}
\end{ruledtabular}
\end{table}

\newpage

\begin{table}[htbp]
Continuation of Table~1.
\bigskip
\begin{ruledtabular}
\begin{tabular}{|c|c|c|}  \hline
1  &2                                     &3                   \\       \hline
22 &   Recoil correction of order $(Z\alpha)^5$      &  -0.2879        \\   \hline
23 &  Recoil correction of order $(Z\alpha)^6$              &  0.0013       \\   \hline
24 & Correction to muon self energy           & -5.0876    \\
   & and muon form factors                      &              \\    \hline
25 &  Radiative-recoil corrections of order           &    -0.0724           \\   
   & $\alpha(Z\alpha)^5$ and proton form factor correction $Z^2\alpha(Z\alpha)^4$ &   \\ \hline
26 &  Nuclear structure correction of order $(Z\alpha)^6$    & -0.0088    \\    \hline
27 &  Contribution of muon form factors $F_1'(0)$, $F_2(0)$    & -0.0112 \\  \hline
28 & VP correction with muon self energy            & -0.0294        \\    \hline
29 & Hadronic vacuun polarization correction             &  0.0830     \\    \hline
30 & Contribution of one-meson exchange  &  -0.1059    \\   \hline
31 & Radiative corrections of order $\alpha(Z\alpha)^6$    & 0.0021    \\    \hline
32 & Summary contribution           &  2249398.5478     \\   \hline
\end{tabular}
\end{ruledtabular}
\end{table}


\begin{thebibliography}{99}
\bibitem{hill}R.~J.~Hill, EPJ Web Conf. {\bf 137}, 01023 (2017).
\bibitem{paz}G.~Paz, e-Print: 1909.08108 [hep-ph]
\bibitem{pineda1}M.~ Horbatsch, E.~A.~Hessels, A.~ Pineda, Phys. Rev. C {\bf 95}, 035203 (2017).
\bibitem{bernauer}J.~C.~Bernauer, EPJ Web Conf. {\bf 234}, 01001 (2020).
\bibitem{carlson}C.~E.~Carlson, Prog. Part. Nucl. Phys. {\bf 82}, 59 (2015).
\bibitem{sgk2019}S.~G.~Karshenboim, V.~G.~Ivanov and S.~I.~Eidelman,
Phys. Part. Nucl. Lett. {\bf 16}, 514 (2019).
\bibitem{crema2010}R.~Pohl et al., Nature {\bf 466}, 213 (2010).
\bibitem{crema2013}A.~Antognini et al., Science {\bf 339}, 417 (2013).
\bibitem{crema}A.~Antognini, F.~Kottmann, F.~Biraben, et al., Ann. Phys. {\bf 331}, 127 (2013).
\bibitem{crema2016}R.~Pohl et al., Science {\bf 353}, 669 (2016).
\bibitem{beyer}A.~Beyer et al., Science {\bf 358}, 79 (2017).
\bibitem{prad}W.~Xiong et al., Nature {\bf 575}, 147 (2019).
\bibitem{bezginov}N.~Bezginov, T.~Valdez, M.~Horbatsch et al., Science {\bf 365}, 1007 (2019).
\bibitem{gilman}R.~Gilman et al. [MUSE Collaboration], arXiv:1709.09753 [physics.ins-det].
\bibitem{fleurbaey}H.~Fleurbaey et al., Phys. Rev. Lett. {\bf 120}, 183001 (2018).
\bibitem{codata}P.~J.~Mohr, D.~B.~Newell, and B.~N.~Taylor, Rev. Mod. Phys. {\bf 88},
035009 (2016).
\bibitem{yerokhin2018}V.~A.~Yerokhin, K.~Pachucki, V.~Patkos, arXiv:1809.00462.
\bibitem{apm2005}A.~P.~Martynenko, Jour. Exp. Theor. Phys.  {\bf 101}, 1021 (2005).
\bibitem{apm2015}A.~A.~Krutov, A.~P.~Martynenko, G.~A.~Martynenko, R.~N.~Faustov, 
Jour. Exp. Theor. Phys. {\bf 120}, 73 (2015).
\bibitem{apm2019}A.~E.~Dorokhov, A.~P.~Martynenko, F.~A.~Martynenko, R.~N.~Faustov, 
Jour. Exp. Theor. Phys. {\bf 129}, 956 (2019).
\bibitem{friar}J.~L.~Friar, Ann. Phys. {\bf 122}, 151 (1979).
\bibitem{borie}E.~Borie, Ann. Phys. {\bf 327}, 733 (2012).
\bibitem{kp1996}K.~Pachucki, Phys. Rev. A {\bf 54}, 1994 (1996).
\bibitem{egs}M.~I.~Eides, H.~Grotch and V.~A.~Shelyuto, Phys. Rep. {\bf 342}, 62 (2001);
M.~I.~Eides, H.~Grotch, and V.~A.~Shelyuto, {\it Theory
of Light Hydrogenic Bound States}, Springer, Berlin-Heidelberg-New York (2007).
\bibitem{cs}G.~K\"allen and A.~Sabry, Mat. Fys. Medd. Dan. Vid. Selesk. {\bf 29}, No. 17, 1 (1955),
in {\it Portrait of Gunnar K\"allen}, edited by C.~Jarlskog, Springer, Switzerland, p.~555 (2014).
\bibitem{kn}T.~Kinoshita and M.~Nio, Phys. Rev. Lett. {\bf 82}, 3240 (1999).
\bibitem{kn1}T.~Kinoshita and M.~Nio, Phys. Rev. D {\bf 60}, 053008 (1999).
\bibitem{kl}T.~Kinoshita and W.~B. Lindquist, Phys. Rev. D {\bf 27}, 853 (1983).
\bibitem{baikov}P.~A.~Baikov and D.~J.~Broadhurst, in {\it New Computing Techniques
in Physics Research IV}, edited by B.~Denby and D.~Perrei-Gallix, World Scientifiс 
Puplishing Co., Singapore, p.~167 (1995).
\bibitem{kataev}R.~N.~Faustov, A.~L.~Kataev, S.~A.~Larin and V.~V.~Starshenko,
Phys. Lett. B {\bf 254}, 241 (1991).
\bibitem{kataev1}D.~J.~Broadhurst, A.~L.~Kataev and O.~V.~Tarasov, Phys. Lett. B {\bf 298}, 445 (1993).
\bibitem{hoang}A.~H.~Hoang, J.~H.~K\"uhn and T.~Teubner, Nucl. Phys. B {\bf 452}, 173 (1995).
\bibitem{hoang1}K.~G.~Chetyrkin et al., Phys. Lett. B {\bf 384}, 233 (1996).
\bibitem{apm2007}A.~P.~Martynenko, Phys. Rev. A {\bf 76}, 012505 (2007).
\bibitem{apm2016}A.~A.~Krutov, A.~P.~Martynenko, F.~A.~Martynenko, and O.S. Sukhorukova,
Phys. Rev. A {\bf 94},  062505 (2016).
\bibitem{hameka}H.~F.~Hameka, Jour. Chem. Phys. {\bf 47}, 2728 (1967).
\bibitem{veselov}M.~G.~Veselov and L.~N.~Labzovsky, Atomic theory: electron shell structure,
M., Nauka (1986).
\bibitem{sgk1996}V.~G.~Ivanov, S.~G.~Karshenboim, Jour. Exp. Theor. Phys.{\bf  82}, 656 (1996).
\bibitem{ikk2009}V.~G.~Ivanov, E.~Yu.~Korzinin and S.~G.~Karshenboim, Phys. Rev. 
D {\bf  80}, 027702 (2009).
\bibitem{kelly}J.~J.~Kelly, Phys. Rev. C {\bf 70}, 068202 (2004).
\bibitem{apm2018}A.~E.~Dorokhov, A.~A.~Krutov, A.~P.~Martynenko, F.~A.~Martynenko,
and O.~S.~Sukhorukova, Phys. Rev. A {\bf 98}, 042501 (2018).
\bibitem{plb2017}R.~N.~Faustov, A.~P.~Martynenko, F.~A.~Martynenko and V.~V.~Sorokin,
Phys. Lett. B {\bf 775}, 79 (2017).
\bibitem{eides1}M.~I.~Eides and H.~Grotch, Phys. Rev. A {\bf 55}, 3351 (1997).
\bibitem{pg}K.~Pachucki and H.~Grotch, Phys. Rev. A {\bf 51}, 1854 (1995).
\bibitem{shabaev}V.~M.~Shabaev, Theor. Math. Phys. {\bf 63}, 588 (1985).
\bibitem{erokhin2015}V.~A.~Yerokhin and V.~M.~Shabaev, Phys. Rev. Lett. {\bf 115}, 233002 (2015).
\bibitem{shabaev2015}V.~A.~Yerokhin and V.~M.~Shabaev, J. Phys. Chem. Ref. Data {\bf 44}, 033103 (2015).
\bibitem{bcr}R.~Barbieri, M.~Caffo and E.~Remiddi, Nuovo Cimento Lett.
{\bf 7}, 60 (1973).
\bibitem{borie1981}E.~Borie, Z. Phys. A {\bf 302}, 187 (1981).
\bibitem{friar2}J.~L.~Friar, J.~Martorell and D.~W.~L.~Sprung, Phys. Rev. {\bf A59},
4061 (1999).
\bibitem{m6}A.~P.~Martynenko, R.~N.~Faustov, Phys. Atom. Nucl. {\bf 64}, 1282 (2001).
\bibitem{sgk2018}E.~Yu.~Korzinin et al., Phys. Rev. A {\bf 98}, 062519 (2018).
\bibitem{korzinin}S.~G.~Karshenboim, E.~Yu.~Korzinin, V.~G.~Ivanov, 
V.~A.~Shelyuto, JETP Lett. {\bf 92}, 8 (2010).
\bibitem{aed}A.~E.~Dorokhov, A.~P.~Martynenko, F.~A.~Martynenko, 
A.~E.~Radzhabov, EPJ Web Conf. {\bf 222}, 03010 (2019).
\bibitem{aed1}A.~E.~Dorokhov, N.~I.~Kochelev, A.~P.~Martynenko, F.~A.~Martynenko, 
R.~N.~Faustov, Phys. Part. Nucl. Lett. {\bf 14}, 857 (2017).
\bibitem{aed2}A.~E.~Dorokhov, N.~I.~Kochelev, A.~P.~Martynenko, F.~A.~Martynenko, 
A.~E.~Radzhabov, Phys. Lett. B {\bf 776}, 105 (2018).
\bibitem{franke}B.~Franke et al., EPJ D {\bf 71}, 341 (2017).
\bibitem{sgk}E.~Yu.~Korzinin et al., Phys. Rev. A {\bf 97}, 012514 (2018).
\bibitem{sgk1}S.~G.~Karshenboim et al., Phys. Rev. A {\bf 81}, 060501 (2010).
\bibitem{apm2006}A.~P.~Martynenko, Phys. Atom. Nucl. {\bf 69}, 1309 (2006).
\end{thebibliography}
\end{document}